\documentclass[12pt,preprint]{aastex}

\slugcomment{Accepted to ApJ, 15 May 2008}

\shorttitle{Infrared Spectroscopic Properties of GV Tau}
\shortauthors{Doppmann et al.}

\newcommand{\teff}{\,T_{\rm eff}}
\newcommand{\logg}{\,\log g}
\newcommand{\vsini}{\,v\sin i}
\newcommand{\rk}{\,r_{K}}
\newcommand{\kms}{\rm \,km\,s^{-1}}
\newcommand{\gsqcm}{\rm \,g\,cm^{-2}}
\newcommand{\Msun}{\,M_\odot}
\newcommand{\Lsun}{\,L_\odot}
\newcommand{\Rsun}{\,R_\odot}
\newcommand{\Mstar}{\,M_*}
\newcommand{\Lstar}{\,L_*}
\newcommand{\Rstar}{\,R_*}
\newcommand{\vlsr}{\,v_{\rm LSR}}
\newcommand{\invcm}{\rm \,cm^{-1}}

\newcommand{\Mdotdisk}{\,\dot M_{\rm d}}
\newcommand{\Msunperyr}{\,M_\odot \rm yr^{-1}}

\begin{document}


\title{Stellar and Circumstellar Properties of the Pre-Main Sequence
Binary GV Tau from Infrared Spectroscopy\altaffilmark{1}}


\author{Greg W. Doppmann\altaffilmark{2},
Joan R. Najita\altaffilmark{2}, and John S. Carr\altaffilmark{3}}

\email{gdoppmann@noao.edu}
\email{najita@noao.edu}
\email{carr@nrl.navy.mil}


\altaffiltext{1}{Data presented herein were obtained at the W.M. Keck Observatory from telescope time allocated to the National Aeronautics and Space Administration through the agency's scientific partnership with the California Institute of Technology and the University of California. The Observatory was made possible by the generous financial support of the W.M. Keck Foundation.}

\altaffiltext{2}{NOAO, 950 North Cherry Avenue, Tucson, AZ 85719}

\altaffiltext{3}{Naval Research Laboratory, Code 7210, Washington, DC 20375}


\begin{abstract}

We report spatially resolved spectroscopy of both components of the
low-mass pre-main-sequence binary GV Tau.  High resolution
spectroscopy in the $K$- and $L$-bands is used to characterize the
stellar properties of the binary and to explore the nature of the
circumstellar environment.  We find that the southern component, GV
Tau S, is a radial velocity variable, possibly as a result of an
unseen low-mass companion.  The strong warm gaseous HCN absorption
reported previously toward GV Tau S \citep{gibb2007} was not present
during the epoch of our observations.  Instead, we detect warm
($\sim$500~K) molecular absorption with similar properties toward the
northern infrared companion, GV Tau N.  At the epoch of our
observations, the absorbing gas toward GV Tau N was approximately at
the radial velocity of the GV Tau molecular envelope, but it was
redshifted with respect to the star by $\sim$13~$\kms$.  One
interpretation of our results is that GV Tau N is also a binary and
that most of the warm molecular absorption arises in a circumbinary
disk viewed close to edge-on.

\end{abstract}



\keywords{infrared: stars -- stars: formation, fundamental parameters,
late-type, low-mass, pre--main sequence --- stars:
circumstellar matter, stars: individual (GV Tau)--- techniques: spectroscopic, radial velocities}



\section{Introduction}

\objectname{GV Tau} (Haro 6-10, Elias 7, HBC 389, IRAS 04263+2426) is an unusual
young stellar object (YSO) in the Taurus molecular cloud.  While the
spectral energy distribution of the system has the strongly rising
2--25~$\micron$ continuum that is characteristic of Class I sources
\citep[e.g.,][]{furlan2007}, millimeter studies find comparatively
weak dust and gas emission from the source compared to other low-mass
embedded YSOs \citep{hogerheijde1998}.  This, and the poorly defined
molecular outflow structure from the source, suggest that the system
lacks a significant envelope component \citep{hogerheijde1998} and may
be a more evolved Class I system.

GV Tau is also a binary \citep[projected separation
1\farcs2;][]{leinert1989}.  It is one of a small population of
low-mass pre-main-sequence binaries in which one component is
optically visible, while the other is optically faint and radiates
primarily at infrared wavelengths.  The optically visible component in
the GV Tau system is the southern component (GV Tau S), a T Tauri
star.  Its stellar properties have been characterized previously based
on high resolution infrared \citep{doppmann2005} and optical
\citep{white2004} spectroscopy.  GV Tau S is unusual in that its
radial velocity has been found to differ significantly (by more than
3-$\sigma$) from both the radial velocity of the surrounding molecular
cloud (by $> 7.5~\kms$) and the radial velocity distribution of other
Taurus sources \citep{doppmann2005,covey2006}.  In addition, $L$- and
$M$-band absorption by gaseous warm organic molecules (HCN,
C$_2$H$_2$, CO) has been reported toward GV Tau S \citep[]{gibb2007},
making it one of the few pre-main-sequence sources in which such
absorption features have been detected.

Less is known about the northern component, the infrared companion of
the binary system (GV Tau N).  It is approximately 100 times fainter
than GV Tau S at optical wavelengths (Stapelfeldt et al., in preparation), is
bright at infrared wavelengths, and experiences significant
extinction, as indicated by the strong water ice and silicate
absorption observed toward the source
\citep{whittet1988,vancleve1994,leinert2001}.  Extended emission (on a
$10\arcsec$ scale) associated with the GV Tau system is found to be
highly polarized (6\%) at infrared wavelengths, a result that is
attributed to scattering by a flattened envelope (i.e., a shell with
an aspect ratio of 10:1) or a disk viewed at high inclination
\citep{menard1993}.  Since strong silicate absorption is found only in
the spectrum of GV Tau N and not GV Tau S, this is taken as evidence
for an edge-on disk associated with GV Tau N \citep{vancleve1994}.

High angular resolution observations provide limited support for this
interpretation.  When studied at $0\farcs08$ resolution at $K$, GV Tau
N is found to be surrounded by a nebula whose shape may be consistent
with the presence of an edge-on disk, but the shape of the nebula is
irregular and varies with time \citep{koresko1999}.  The irregular
structure is in marked contrast to the symmetric lens-like shape of
scattered light distribution seen in the HK Tau B system, a system
that is also believed to possess a nearly edge-on disk
\citep{stapelfeldt1998, koresko1998}.  Adding further complexity,
significant photometric variability is found for both components at
$1.6-4.5~\micron$ on timescales as short as a month
\citep{leinert2001}, with the $3.1~\micron$ water ice band absorption
strength also varying toward both components.

Here we use spatially resolved high resolution infrared spectroscopy
of both binary components to characterize the stellar and
circumstellar components of the GV Tau system.  Our $K$- and $L$-band
spectroscopic observations are described in \S~2 and the detected
spectral features are described in \S~3.  We use these data to
characterize the stellar and circumstellar properties of the system
(\S~4).  The results are discussed in \S~5 and our conclusions
presented in \S~6.

\section{Observations and Data Reduction}

\subsection{Spectroscopic Observations}

$K$- and $L$-band infrared spectra were obtained on 2007 January 03 using
the cryogenic echelle facility spectrometer, NIRSPEC
\citep{mclean1998}, on the 10-m Keck II telescope atop Mauna Kea,
Hawaii. Spectra were acquired through the 0\farcs432 (3 pixel) wide
slit, providing spectroscopic resolution $R \equiv \lambda / \delta
\lambda = 24,000$ (12.5~$\kms$).  The echelle and cross disperser
angles were oriented to obtain $L$-band spectra of transitions of HCN
and C$_2$H$_2$ (e.g. 3.04 - 3.09~$\micron$, order 25).  We used the KL
blocking filter to image the $L$-band orders onto the instrument's
1024 $\times$ 1024 InSb detector.  Immediately following the $L$-band
observations, the echelle and cross disperser gratings were
re-oriented to allow key $K$-band features (i.e., Mg I, Al I, Na I,
$\nu$=2-0 CO, and Br$\gamma$) to fall onto the detector.  In this
configuration, orders 32 -- 38 (non-contiguous) were imaged through
the NIRSPEC-7 blocking filter onto the detector array.

Both binary components of GV Tau were acquired in the slit
simultaneously, at a position angle on the sky of 356$\arcdeg$
\citep{leinert1989}.  Initially, GV Tau N (the brighter component in
our observations of $\lambda >2.09~\micron$) was centered in the slit.
During the $L$- and $K$-band observations the slit was held physically
stationary to avoid slight grating angle shifts caused by vibrations
from the slit rotator motor.  During the time we integrated on the
binary pair ($<$25 min. total), the measured flux ratio of the
components in the slit did not change systematically.  This is
consistent with the negligible expected motion of the binary pair
within the slit as the non-equatorially mounted telescope tracked.
Despite some wind shake in the direction across the slit in the
east-west direction, the individual components (separated by
1$\farcs$2) were well resolved in median seeing ($\sim$0$\farcs6$),
allowing separate spectral extractions of the southern and northern
components.

The data were acquired in pairs of exposures of 40s and 60s in the
$L$- and $K$-bands, respectively.  The telescope was nodded
$\pm6\arcsec$ from the slit center in an ABBA sequence along the
24$\arcsec$ long slit so that each frame pair would contain object and sky
in both nod positions.  To correct for telluric absorption, we also
obtained spectra of an early-type star located close to GV Tau in the
sky (HR 1412), before and after the GV Tau observations ($L$- and
$K$-band settings, respectively).  Spectra of the internal NIRSPEC
continuum lamp were taken for flat fields at the $K$- and $L$-band
settings.  Exposures of the Argon, Neon, Krypton, and Xenon Arc lamps
provided wavelength calibration for all the $K$-band orders, except in
order 33 where we used the telluric lines for calibration due to the
scarcity of arc lamp lines in this order.  We also obtained short
exposure (10 $\times$ 0.1s) $K$- and $L$-band images of the GV Tau
binary using the slit viewing camera detector (SCAM) in order to
measure the relative brightness of the two components.

\subsection{Data Reduction}

The spectra were reduced using standard IRAF packages
\citep{massey1992,massey1997}.  Sky subtracted beam pairs were divided
by a normalized flatfield.  Individual echelle orders were parsed from
the multi-order, co-added images.  Bad pixels (i.e., hot or low
responsivity pixels) in each order were identified by inspection, and
removed by interpolation using $\it{fixpix}$ in IRAF.  To better
remove telluric emission features which are more severe in the
$L$-band, we rectified the slit in our $L$-band data using 3rd order
transform solution derived from the brightest emission sky lines
(typically 10-12 per order) traced along the slit length at each beam
position before sky subtraction.  Transformed and cleaned orders were
extracted using the $\it{apall}$ task.  Extractions for each binary
component of GV Tau were based on a signal profile down to 50\% (FWHM)
on both sides of the profile peak, to minimize the contaminating flux
from the nearby companion.  For the $L$-band spectra, residual
background sky was subtracted in the extraction using selected regions
of the profile cut that were well outside of the profile of the
spatially double-peaked binary.

Wavelength calibration in the $L$-band was achieved using selected
telluric absorption features present in the spectra of the telluric
standard star (HR 1412).  Stronger unblended telluric lines were
selected over weaker ones, while avoiding very saturated lines whose
depths went to zero.  Rest wavelengths for the absorption features
were obtained from the HITRAN database \citep{rothman1998}.  Most or
all of these same lines were used to derive a wavelength solution for
the target, with the exception of those lines that might fall very
close in wavelength to emission or absorption lines present in the
target spectrum.

Telluric features present in the spectra of GV Tau at each nod
position in the slit were removed by dividing by the spectrum of the
telluric standard (HR 1412) observed at the same nod position.
Several weak stellar absorption lines were present in the K-band
orders of the standard due to its relatively late spectral type (A7
III). These atomic lines were modeled and removed by dividing the
standard spectrum by the best fit stellar synthetic spectrum before
division into the GV Tau spectrum.  The stellar synthetic spectra were
generated with the program MOOG \citep{sneden1973} using the NextGen
model atmospheres \citep{hauschildt1999}.  The initial line list was
taken from \citet{kurucz1993}, and individual line parameters were
adjusted to fit the observed disk-center solar spectrum of
\citet{livingston1991}. For HR 1412, we used a model with $\teff =
7600$ K and $\logg = 3.5$.  Small adjustments were made in the
elemental abundances to give the best empirical fit to HR 1412.  We
measured $\vsini = 75 \kms$, in agreement with published values
\citep{royer2002}.

Br $\gamma$ absorption at 2.166 $\micron$ was also present in the
telluric standard.  We therefore used the synthetic telluric spectral
modeling program, AT (Grossman, private communication) to model the
telluric features that we observed in this region of the HR 1412
spectrum.  We fit the depths of the observed telluric lines and then
used the best-fit synthetic spectrum to divide out the telluric lines
in the GV Tau spectra in this order (order 35,
Fig. \ref{fig-kband.ord33}c).

In order to view the true continuum shape within each echelle order,
we then multiplied the resulting GV Tau spectra by the spectral slope
of a 7600~K blackbody (Figs. \ref{fig-kband.ord33}a --
\ref{fig-kband.ord33}e).

Simple aperture photometry of the SCAM images showed that GV Tau N was
35-40\% brighter than GV Tau S in the K+open, NIRSPEC7, and KL
filters.  The relative flux of the GV Tau N and S $K$-band spectra in
orders 33 and 34 (Figs. \ref{fig-kband.ord33}a \&
\ref{fig-kband.ord33}b), which are located in the middle of the
$K$-band filters above, are consistent with the aperture photometry.

\section{Detected Spectral Features}\label{sec-k+lspec}

\begin{figure}
\plotone{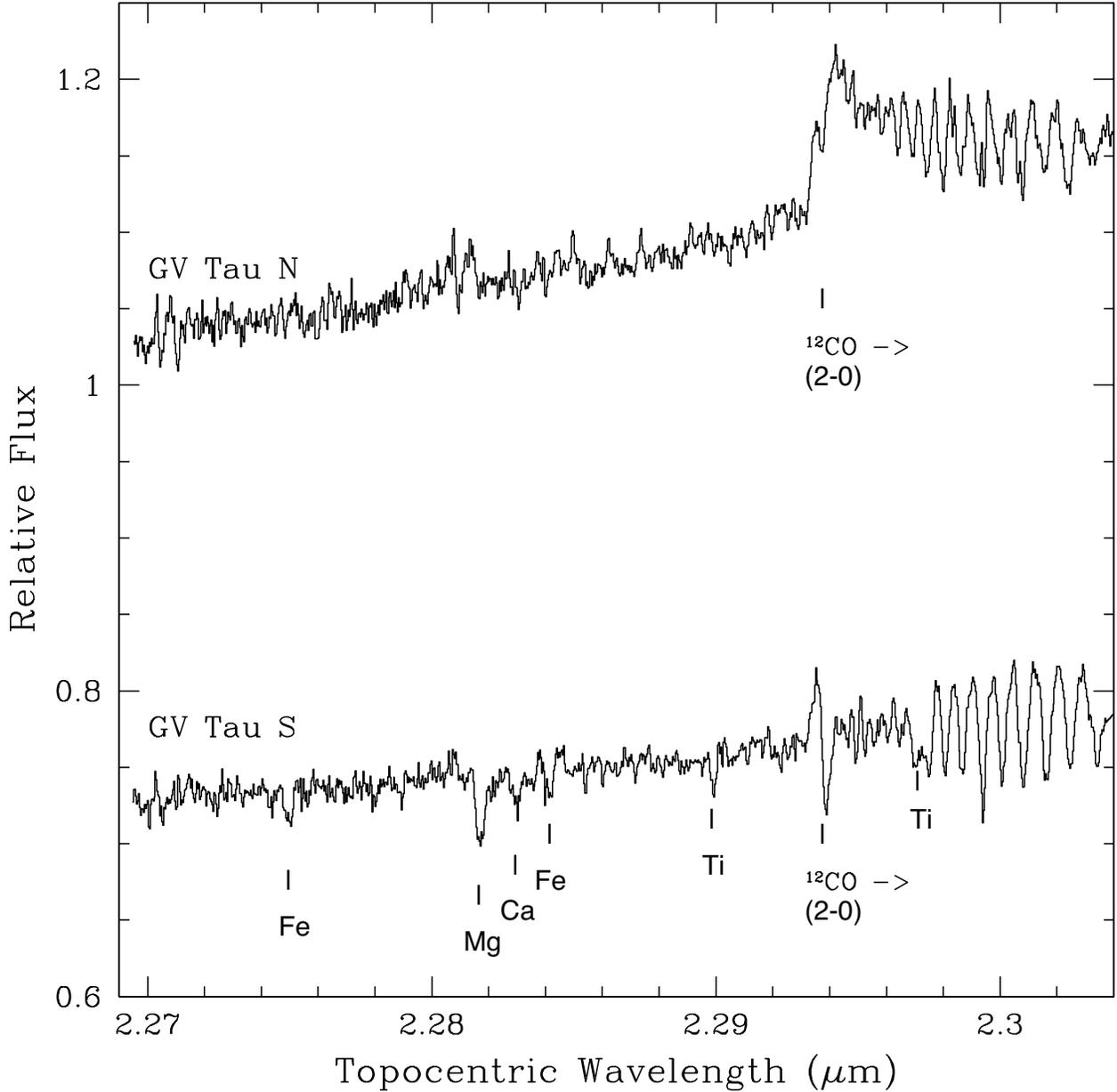}
\caption
{\label{fig-kband.ord33} 
Simultaneous spectra of GV Tau N (top) and GV Tau S (bottom) in the
$K$-band in order 33 (Fig. \ref{fig-kband.ord33}a), order 34
(Fig. \ref{fig-kband.ord33}b), order 35 (Fig. \ref{fig-kband.ord33}c),
order 36 (Fig. \ref{fig-kband.ord33}d), and in the $L$-band order 25
(Fig. \ref{fig-kband.ord33}e).  The relative flux ratio (f$_{\rm
{South}}$/f$_{\rm {North}}$) of the two components is $\sim$0.7 (order
33, Fig. \ref{fig-kband.ord33}a), 0.74 (order 34,
Fig. \ref{fig-kband.ord33}b), 0.85 (order 35,
Fig. \ref{fig-kband.ord33}c), $\sim$1.0 (order 36,
Fig. \ref{fig-kband.ord33}d), and 0.45 (order 25,
Fig. \ref{fig-kband.ord33}e).  The $K$-band orders show emission
features in both objects, while absorption lines dominate the spectra
in GV Tau S.  The $L$-band spectrum of GV Tau N
(Fig. \ref{fig-kband.ord33}e) shows strong P-branch absorption lines of
HCN (red tick marks, labeled) and several weaker C$_2$H$_2$ lines
(blue tick marks), while these features are absent in GV Tau S.
Spectral regions with poor telluric transmission have been excised
from Fig. \ref{fig-kband.ord33}e.
}
\end{figure}

Emission and absorption features in GV Tau N and GV Tau S were
detected in the $K$-band spectra of each object
(Figs. \ref{fig-kband.ord33}a -- \ref{fig-kband.ord33}d).  In order 33
(2.269 -- 2.304~$\micron$, Fig. \ref{fig-kband.ord33}a), $\nu$=2-0 CO
overtone emission and absorption are present in both GV Tau N and GV
Tau S.  The emission component dominates the spectrum in GV Tau N in
this order, while the absorption dominates in GV Tau S.  Lines due to
absorption of neutral atomic species that are present in GV Tau S in
this order (e.g., Fe I at 2.2747~$\micron$ and Mg I at
2.2814~$\micron$) are absent in GV Tau N.  The $K$-band absorption
features in both components are presumed to originate from the stellar
photosphere.

Emission features of Na I and H$_2$ dominate the structure of the GV
Tau N spectrum in order 34 (2.203 -- 2.236~$\micron$,
Fig. \ref{fig-kband.ord33}b).  Interestingly, the broad emission lines
near 2.2065 and 2.2090~$\micron$ in GV Tau N appear to show absorption
components possibly of a photospheric origin.  The S(0) 1-0 H$_2$
emission at 2.2235~$\micron$ is detected in both components, and is
spatially extended (see $\S$~\ref{sec-h2.emis}).  Other than the H$_2$
emission, the spectrum of GV Tau S in this order displays neutral
atomic absorption lines of photospheric origin
(Fig. \ref{fig-kband.ord33}b).

The spectra in order 35 (2.140 -- 2.173~$\micron$,
Fig. \ref{fig-kband.ord33}c) are characterized by strong H I Br$\gamma$
emission in both GV Tau N and GV Tau S.  The continuum level near the
Brackett $\gamma$ line is $\sim$20\% higher in GV Tau N, while the
equivalent width of the emission line is 3 times greater in GV Tau S.

In order 36 (2.081 -- 2.113~$\micron$, Fig. \ref{fig-kband.ord33}d),
the spectrum of GV Tau N appears featureless.  In contrast, GV Tau S
exhibits photospheric absorption lines (e.g., Mg I at 2.1065~$\micron$
and Al I at 2.1099~$\micron$), which are useful as diagnostics of the
stellar effective temperature and surface gravity.

With our $L$-band spectra in order 25 (3.0450--3.0865~$\micron$,
Fig. \ref{fig-kband.ord33}e), we detect strong HCN absorption
($\sim$10\% deep) and weaker C$_2$H$_2$ absorption in GV Tau N.  In
contrast, no molecular absorption is detected in the GV Tau S
spectrum.  In \S~\ref{sec-molecular.abs} we characterize the molecular
absorption in GV Tau N.

\section{Results}

\subsection{Stellar  and Circumstellar Properties}\label{sec-gvt.stellprops}
\subsubsection{South Component}\label{sec-gvts.stellprops}

\begin{figure}
\plotone{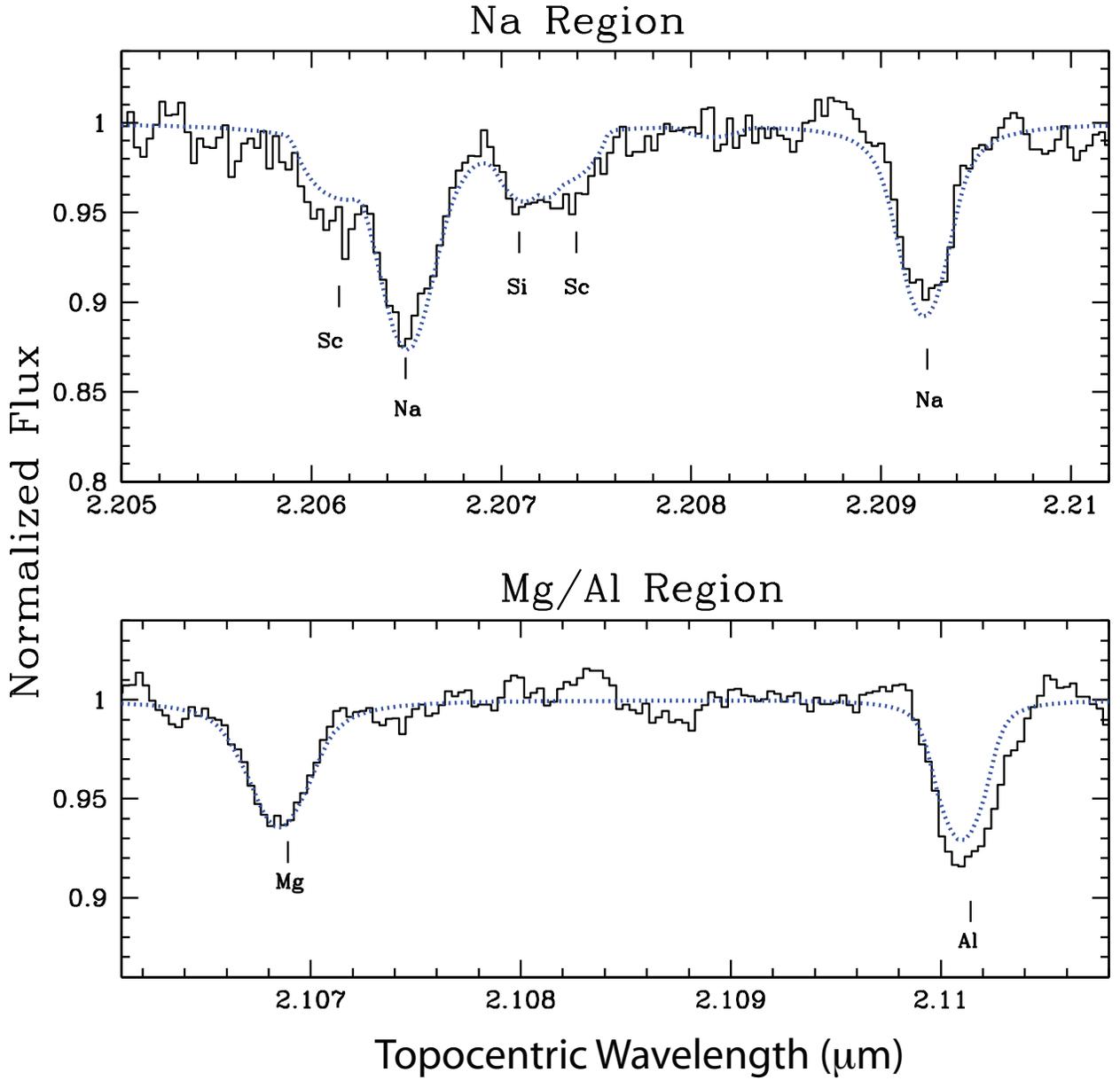}
\caption
{\label{fig-kband.gvts.modelfit}
Two regions of the $K$-band of GV Tau S (black histogram) that show
stellar photospheric absorption lines. A spectral synthesis model
(dotted blue) with $\teff$=3800~K, $\logg$ = 4.0, $\vsini$ = 24~$\kms$
(includes instrumental broadening), along with a veiling of $\rk$ =
2.5 fits the spectrum.  The model fit in the top panel is at a radial
velocity of $\vlsr$ = 9.1 $\kms$ (see Table 1).}
\end{figure}

All four $K$-band orders of the GV Tau S spectrum display stellar
photospheric absorption by neutral atomic features (e.g., Na I, Si I,
Sc I, Mg I, Al I) and $\nu$=2-0 CO.  We used synthetic stellar
spectral models generated by the synthesis program MOOG
\citep{sneden1973} and using NextGen stellar atmosphere model
structures \citep{hauschildt1999} to constrain the stellar properties
of GV Tau S.  Our modeling focused on three spectral sub-regions: The
``Na" region (2.204--2.210~$\micron$), the ``Mg/Al" region
(2.104--2.111~$\micron$), and the ``CO" region
(2.292--2.300~$\micron$).  These regions of the $K$-band contain the
strongest absorption features with which to measure the stellar
properties of low-mass YSOs \citep{doppmann2005}.  The absorption
lines of Na I and Mg I are particularly gravity and temperature
sensitive, but in the opposite sense from each other.  For example, at
cool effective temperatures (3200--4500~K) and subdwarf surface
gravities ($3.5 \le \logg \le 4.5$) Na and Mg lines both grow stronger
as $\logg$ increases, but an increase in $\teff$ causes Mg lines to
grow while the Na lines weaken.  A simultaneous fit of synthesis
models to the Na and Mg/Al regions
(Fig. \ref{fig-kband.gvts.modelfit}) constrains these key stellar
properties in reducing or breaking the temperature-gravity degeneracy
displayed in a single absorbing species \citep{doppmann2005}.

The best model fit to these wavelength sub-regions yields $\teff$=3800
$\pm$100~K, $\logg$=4.0 $\pm$0.17, $\vsini$ rotation= 24
$\pm$3~$\kms$, and $K$-band veiling $\rk$=2.5 $\pm$0.2
(Fig. \ref{fig-kband.gvts.modelfit}).  From the $\teff$ and $\logg$
derived above, we estimate a stellar luminosity, mass, and radius of
$\Lstar=0.3 \Lsun, \Mstar=0.5 \Msun, \Rstar=1.2 \Rsun$ using the
pre-main-sequence model tracks of \citet{siess2000}, where a 3 Myr
isochrone was consistent with our values of temperature and gravity.
The derived luminosity and effective temperature imply an estimated
$K$-band extinction toward GV Tau S of $A_{\rm K}=0.7$ based on the
most recent photometery from the literature \citep[2000 March
07;][]{leinert2001}, which reported a value ($m_{\rm K} = 8.61$)
that was within 1-$\sigma$ of the average in the study over the time
period from September 1988 to March 2000 ($m_{\rm K} = 8.1 \pm 0.7$).
Our derived $K$-band extinction is consistent with the average optical
extinction of $A_{\rm V} = 5.6$ estimated by \citet{koresko1997}.

Our derived $\teff$, $\vsini$, and $\vlsr$ values are consistent with
the high resolution optical study of \citet{white2004} given the low
signal to noise (S/N $<$ 10) in their 1999 December 06 observation of GV
Tau S.  The moderately high veiling we derive ($\rk=2.5$) is more
characteristic of Class I or flat spectrum YSOs than of T Tauri stars,
although still within the range of veiling values measured in a sample
of Ophiuchus Class II sources \citep{doppmann2003}.

These results agree well with the veiling measured from the previous
observation of GV Tau S from 2001 November 06 \citep{doppmann2005}.
\citet{doppmann2005} quote a veiling value of $\rk$= 1.8, which
includes a correction for a systematic effect seen in the best fit
synthesis models to observations of MK standards \citep[see Eqn. 1 in
\S~3.6 of][]{doppmann2005}.  The measured veiling without the
correction for the systematic effect was $\rk$= 2.4, which is
consistent with our measured value.

The constraint on effective temperature and surface gravity is
consistent with the results from \citet{doppmann2005}.
\citet{doppmann2005} reported a somewhat warmer effective temperature
($\teff=4500K$) and higher surface gravity ($\logg=4.4$), which was
based on fits to synthetic spectra of a smaller number of photospheric
absorption features (excluding Sc I lines at 2.2058 and 2.2071 $\mu$m)
at similar signal-to-noise and slightly lower spectral resolution
($R$=18,000).  A comparably good fit to the 2001 data can be obtained
with $\teff=4000K$ and $\logg=3.9$ that includes the Sc I lines above
as was done in this study.

The radial velocity of GV Tau S was measured by shifting a synthetic
template spectrum to independently match each of 12 selected
absorption lines that were present in orders 33 and 34 (see Table 1).
The error was determined from the quadrature sum of the systematic
error, dominated by the uncertainty in the wavelength calibration of
the data, and the statisical error, measured from the variation in the
radial velocity values among the individual absorption lines.  Our
measured $\vlsr$ of GV Tau S differs from previous measurements in the
literature for this source.  In 2001, the radial velocity of GV Tau S
was measured to be $\vlsr$= --6.2 $\pm 1.5~\kms$ \citep{covey2006},
blueshifted by $\sim$9--15~$\kms$ relative to the observations of this
study and that of \citet[][$\vlsr=+3.1 \pm 3.8~\kms$]{white2004}.  As
a result, this source was flagged as a radial velocity outlier, since
its radial velocity was $>$~3-$\sigma$ from the mean of all the Taurus
sources in the survey \citep{covey2006}.  Our 2007 observations now
place the radial velocity of GV Tau S ($\vlsr= 9.4 \pm1.7~\kms$) close
to the mean $\vlsr$ ($4.6 \pm 1.8~\kms$) of the sources in the 2001
Taurus survey \citep{doppmann2005,covey2006}.  We discuss the radial
velocity variations further in \S~\ref{sec-gvts.binary}.

\begin{figure}
\plotone{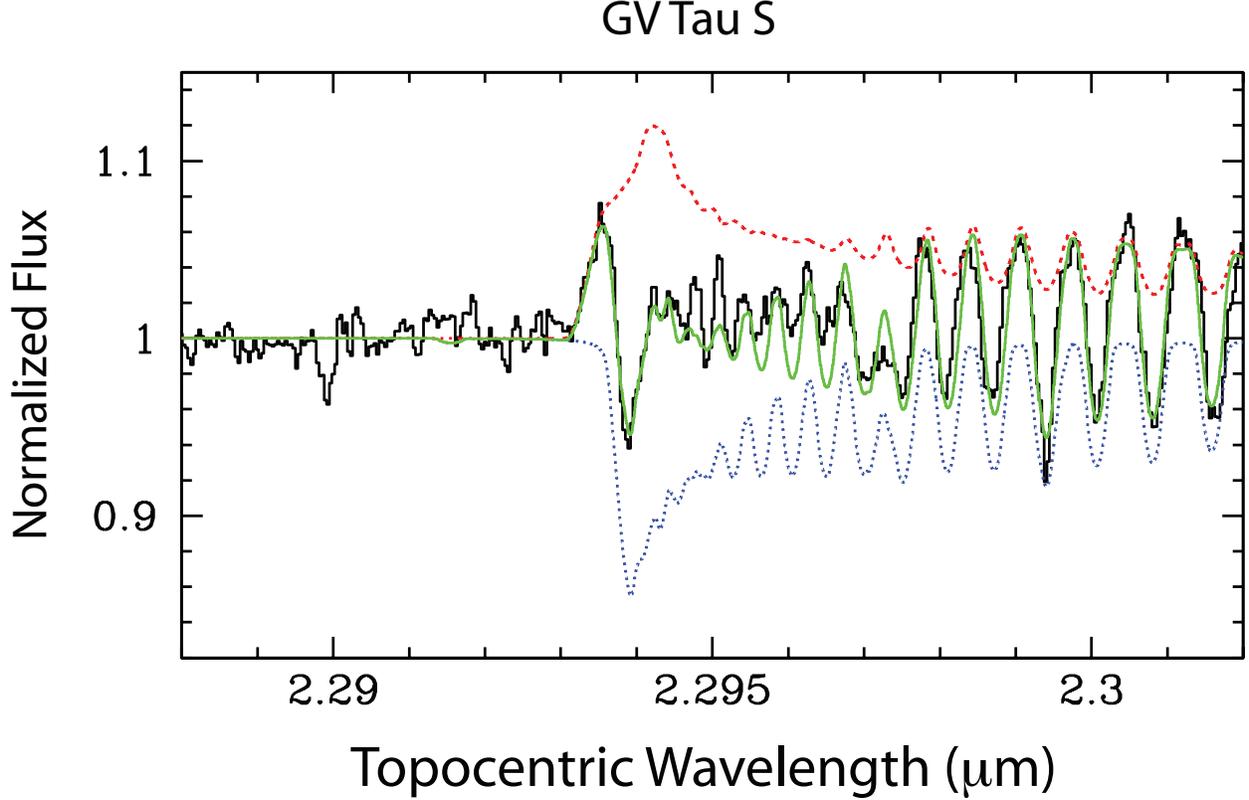}
\caption
{\label{fig-kband.gvts.modelco}
The CO overtone spectral region of GV Tau S (black histogram) showing
both emission and absorption components.  We fit the observed spectrum
using a model of disk emission and stellar photospheric absorption
components at the same radial velocity.  Our disk emission model
(dashed red line) has the properties: $\vsini$~(inner~edge) =
88~$\kms$, $R_o/R_i$~=~7, $T = 3500~(r/R_i)^{-0.5}$~K, $\Sigma =
0.5~(r/R_i)^{-0.5}\gsqcm$, where $r$ is the disk radius.  The stellar
model (dotted blue line) has a temperature $\teff$=3800~K, gravity
$\logg$ =4.0, rotation $\vsini=24~\kms$, and veiling $\rk =2.6$.  The
radial velocity of the model fit ($\vlsr= 9.4~\kms$) is consistent
with the stellar radial velocity of GV Tau S reported in Table 4.  The
stellar model used in the fit is consistent with the fit to the Na and
Mg/Al regions (Fig. \ref{fig-kband.gvts.modelfit}).  The combined disk
emission and photospheric absorption model (solid green) is a
reasonable fit to the observed spectrum.}
\end{figure}

As described in $\S$~\ref{sec-k+lspec}, the CO $\nu$=2-0 bandhead of
GV Tau S has both emission and absorption components.  Using the best
fit stellar parameters obtained from the Na and Mg/Al regions of the
GV Tau S spectrum as a constraint on the stellar photospheric
properties, we fit the CO feature with a composite model of
circumstellar disk emission and stellar photospheric absorption
(Fig. \ref{fig-kband.gvts.modelco}).  The emission component is
modeled as arising from a differentially rotating disk that is in
chemical equilibrium.  The emission arises between an inner radius
$R_i$, at which the projected disk rotational velocity has a specified
value of $\vsini$, and an outer radius $R_o$.  The disk temperature
and column density are modeled as simple radial gradients that
decrease as power laws \citep{carr1993,carr2004}.  The star and disk
are placed at the same radial velocity and the composite model
spectrum is smoothed to the $13\kms$ velocity resolution of NIRSPEC.
The model fit parameters are given in the caption to
Fig. \ref{fig-kband.gvts.modelco}.

\subsubsection{North Component}\label{sec-gvtn.stellprops}

\begin{figure}
\plotone{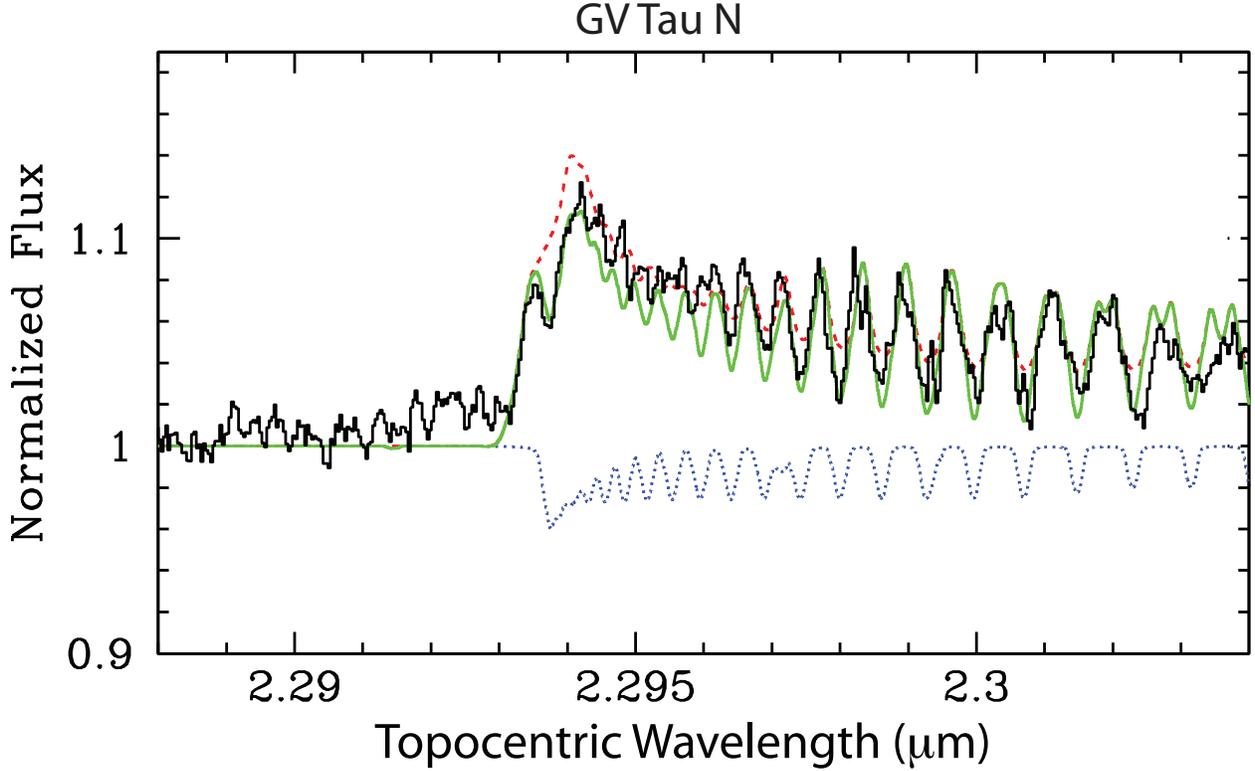}
\caption
{\label{fig-kband.gvtn.modelco}
The observed $\nu$=2-0 CO spectrum of GV Tau N (black histogram) also
showing both emission and absorption components.  A model of CO emission
from a rotating disk (dashed red line) is constructed from the
following parameters: $\vsini$~(inner~edge) = 95~$\kms$, $R_o/R_i$~=~7, $T
= 3000~(r/R_i)^{-0.35}$~K,~$\Sigma = 2.0~(r/R_i)^{-0.5}\gsqcm$, where
$r$ is the disk radius.  The disk model is combined with a model of
stellar photospheric CO absorption (dotted blue line) at the same
radial velocity having the physical parameters $\teff$ = 4100~K,
$\logg$ =4.0, $\vsini$ =15~$\kms$, and $\rk$ = 12.0.  At
$\vlsr=-4.5\kms$ (Table 4), the combination (solid green), is a
reasonable fit to the observed CO feature in GV Tau N.}
\end{figure}

\begin{figure}
\plotone{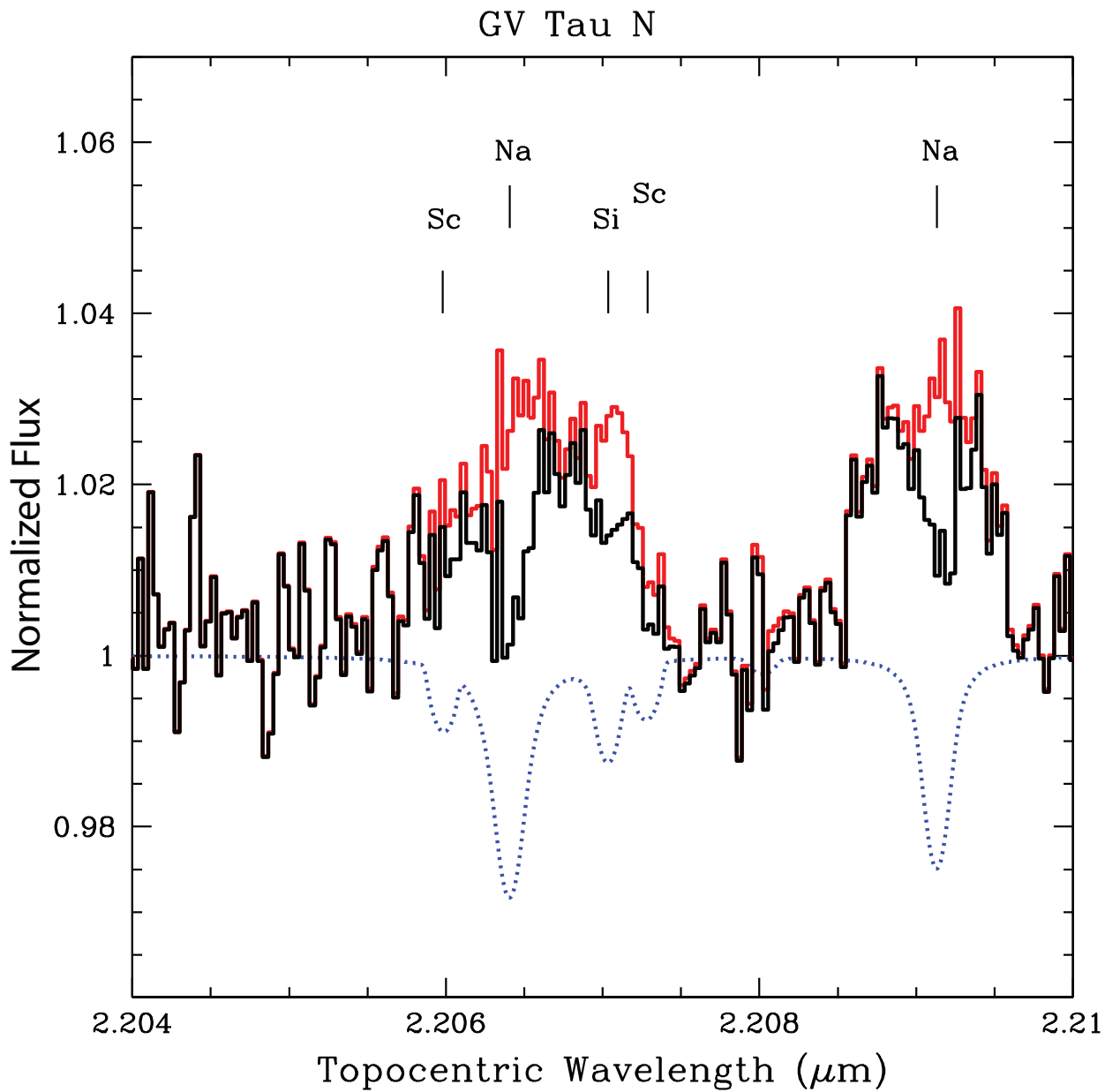}
\caption
{\label{fig-kband.gvtn.modelna}
High resolution (R = 24,000) spectrum of GV Tau N (black histogram)
with a stellar photospheric model fit (dotted blue) to the absorption
component that appears to be present within the broad emission
observed in the Na region of the $K$-band. The Na absorption component
is reasonably fit with the same stellar parameters used in modeling
the CO region (Fig. $\ref{fig-kband.gvtn.modelco}$; $\teff$=4100~K,
$\logg$=4.0, $\vsini=15\kms$, and $\vlsr=-4.5\kms$) and with veiling
$\rk=15$.  The emission spectrum (red histogram), obtained by
subtracting the model absorption from the observed spectrum, reveals
broad emission in the Na lines and possibly the Si line.}
\end{figure}

The spectrum of GV Tau N shows emission and absorption components in
both the CO and Na regions (Figs. \ref{fig-kband.gvtn.modelco} \&
\ref{fig-kband.gvtn.modelna}) of the $K$-band.  We attribute the
emission to a circumstellar disk; the absorption features are presumed
to arise from the stellar photosphere, implying a late spectral type
for the star.  We assume a stellar mass of $M_*$=0.8$\Msun$ and an age
of $\tau$=3 Myr for this object, consistent with the typical values
found for low-mass YSOs in Taurus \citep{briceno2002}.  
These assumptions imply an effective temperature of $\teff$=4100~K and a
surface gravity of $\logg \sim4.0$ from pre-main-sequence evolutionary
model tracks \citep{siess2000}.

Using these assumed stellar properties, we modeled the CO feature
observed in GV Tau N as a combination of circumstellar disk emission
and stellar photospheric absorption.  As in the case of GV Tau S, the
emission component is fit with a simple model of CO emission from a
differentially rotating disk.  The details of the model fit are given
in the caption to Fig. \ref{fig-kband.gvtn.modelco}.  We find a good
fit to the observed CO overtone emission
(Fig. \ref{fig-kband.gvtn.modelco}) with the additional assumptions of
slow stellar rotation ($\vsini$= 15~$\kms$) and heavy veiling
($\rk$=12).  Such a substantial infrared veiling greatly exceeds what
has been measured from near--IR photospheric absorption lines in past
studies \citep[e.g., $\rk \le
4.5$,][]{luhman1998,luhman1999,doppmann2003,doppmann2005}.  It also
results in large uncertainties in the inferred $K$-band extinction.

Broad Na emission has been observed in several Class I and flat
spectrum protostars, always accompanied by $\nu$=2-0 CO emission
\citep{doppmann2005}.  The absorption lines in the Na region, apparent
within the broader emission features, can be fit with the same stellar
photospheric model and somewhat larger veiling ($\rk=15$,
Fig. \ref{fig-kband.gvtn.modelna}) than was used in fitting the CO
region.  If we subtract the stellar photospheric model from the
observed spectrum, the resulting spectrum is similar to the Na
emission seen in other T Tauri stars.  The absence of detected
spectral features in the Mg/Al region of GV Tau N in our data
(signal-to-noise $\sim170$) is also consistent with the model
assumptions.

The radial velocities of the CO and Na absorption components in GV Tau
N are in agreement with one another ($\vlsr=-4.5\kms$, see Table 2),
and with the radial velocity of the emission component from which they
arise.  Given the assumed mass $\Mstar=0.8 \Msun$, we estimate the
stellar luminosity, radius, and $K$-band extinction (i.e., $\Lstar=0.6
\Lsun,~\Rstar=1.4 \Rsun$, and $A_K=2.6$) from the photospheric
absorption we find in our spectra, using \citet{siess2000}
evolutionary model tracks and \cite{leinert2001} photometry from 2000
March 07 ($m_{\rm K}=8.66$).

The unknown apparent magnitude of GV Tau N on the date
of our observations introduces an uncertainty in the derived $K$-band
extinction and veiling.  If the apparent magnitude was the historical
average of $m_{\rm K}=9.8$ \citep{leinert2001}, instead of the value
reported on 2000 March 07 of $m_{\rm K}=8.66$ \citep{leinert2001},
then the extinction ($A_K$) would increase by 1.2.  If we had assumed
a higher ($\teff$=4300~K) or lower ($\teff$=3900~K) temperature than
the value we use here ($\teff$=4100~K), which was based on the IMF
peak \citep{briceno2002}, this would imply a veiling ($\rk$) of 11 or
13, and an extinction ($\Delta A_{\rm K}$) of +0.5 or -0.6 relative to
the derived value presented above ($A_{\rm K}$=2.6).

\subsection{Brackett $\gamma$ Emission}\label{sec-brg.emis}

\begin{figure}
\plotone{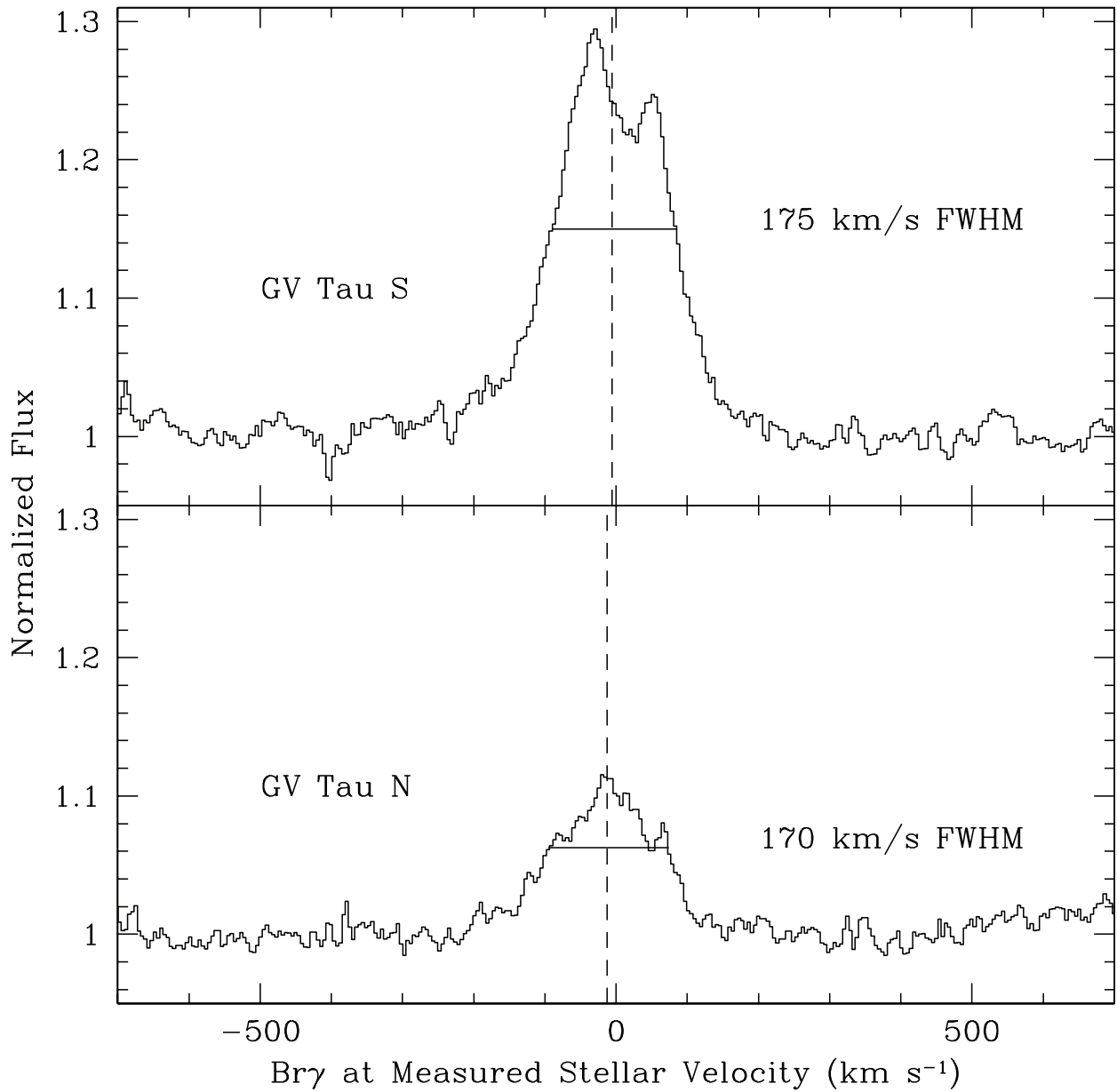}
\caption
{\label{fig-kband.brg}
Resolved Br$\gamma$ emission profiles for GV Tau S (top), and GV Tau N
(bottom).  Zero velocity is defined as the stellar velocity, traced by
the radial velocity of $K$-band absorption lines (Table 4).  The
vertical dashed lines show the respective velocity centroids of the
Br$\gamma$ feature, while the solid horizontal lines illustrate the
feature width (FWHM).  The velocity centroid relative to the stellar
velocity is blueshifted by 5.7 and 12.5~$\kms$ for GV Tau S and N,
respectively (Table 1 and 2).}
\end{figure}

We detect H I Br$\gamma$ emission in both components of GV Tau
(Fig. \ref{fig-kband.brg}).  The broad emission profiles (i.e., 175
and 170~$\kms$ FWHM, GV Tau S and N, respectively) are consistent with
the Br$\gamma$ width of other active T Tauri stars
\citep{najita1996,folha2001}.  In GV Tau S, where the equivalent width
is three times greater than in GV Tau N (EW$_{\rm South} = -3.9$ \AA
), the centroid velocity is very close to the stellar velocity, as
measured by the stellar photospheric absorption lines (see
$\S$~\ref{sec-gvts.stellprops}).  However, the centroid velocity of
the Br$\gamma$ line in GV Tau N is blueshifted by 12.5~$\kms$ relative
to the CO absorption component ($\S$~\ref{sec-gvtn.stellprops}).
Blueshifted Br$\gamma$ centroids, which are common among T Tauri
stars, are consistent with an origin for the emission in gas infalling
in a stellar magnetosphere
\citep{najita1996,muzerolle1998b,folha2001}.

We can use the Br$\gamma$ line strength to place a rough constraint on
the contribution of stellar accretion in GV Tau N and GV Tau S to the
bolometric luminosity of the system.  We first convert the emission
equivalent width to a line luminosity using the 2000 March 07 $K$-band
photometry from \citet{leinert2001}, and the estimated $K$-band
extinction for each component (\S~\ref{sec-gvtn.stellprops} and
\S~\ref{sec-gvts.stellprops}).  The estimated Br$\gamma$ luminosities
for GV Tau N and S are $9\times 10^{-5} \Lsun$ and $5\times 10^{-5}
\Lsun$, respectively.  Using the empirical relation given by
\citet{muzerolle1998b} in their Figure 4, we infer accretion
luminosities of 0.1 and 0.2 $\Lsun$ for GV Tau N and S, respectively.
Thus, we estimate the total luminosity (stellar+hot accretion) of the
GV Tau system to be 1.2 $\Lsun$, below what has been estimated for its
bolometric system luminosity \citep[$L_{\rm bol}=7 - 9
\Lsun$;][]{kenyon1995,furlan2007}.

\subsection{H$_2$ emission}\label{sec-h2.emis}

\begin{figure}
\epsscale{0.80}
\plotone{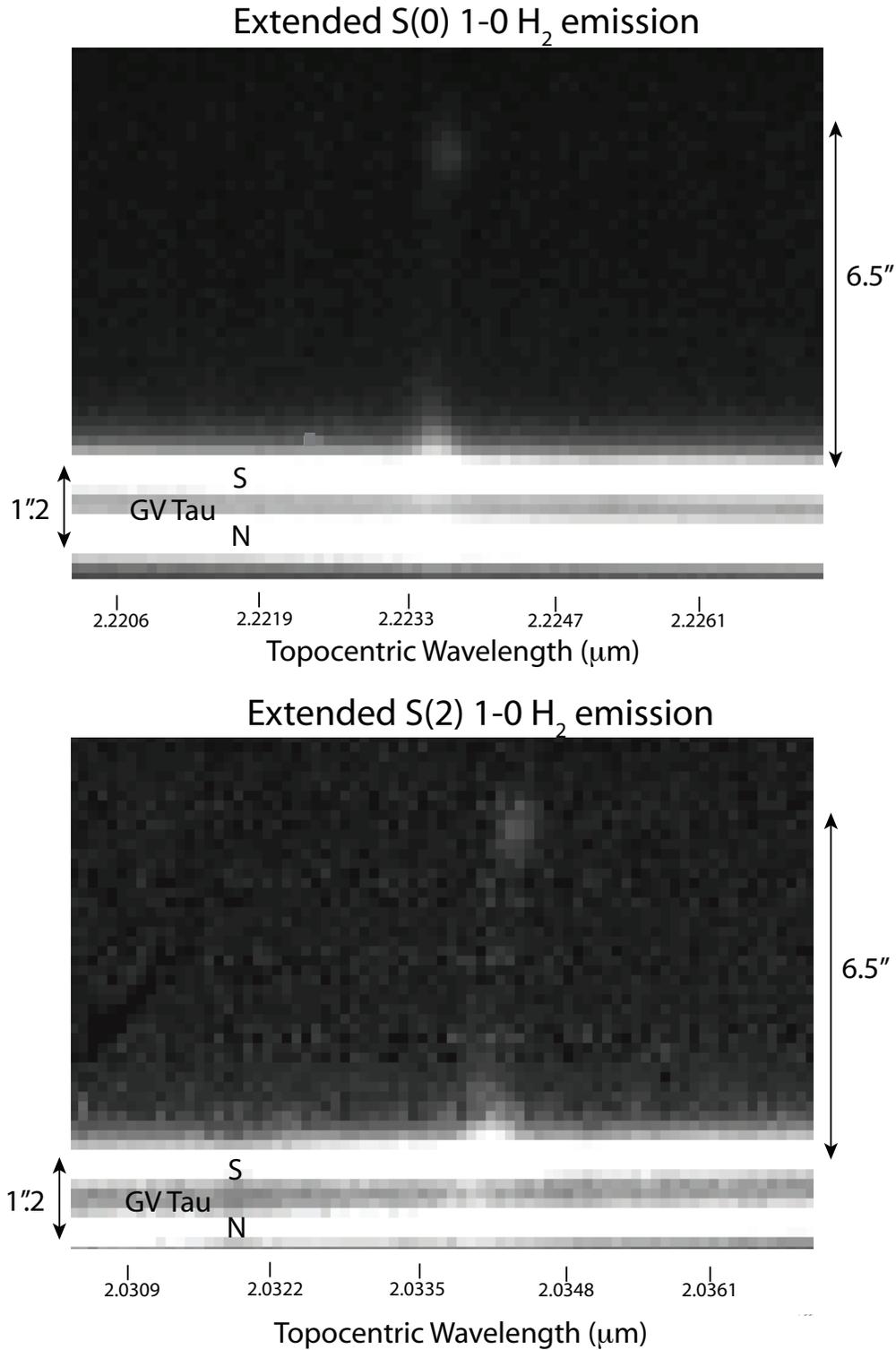}
\caption
{\label{fig-gvtau.h2emis}
Spectral images of both components of GV Tau showing extended H$_2$
emission.  The S(0) 1-0 emission (2.2233~$\micron$, top) and the S(2)
1-0 emission (2.0338~$\micron$, bottom) are observed at the positions
of both GV Tau S and GV Tau N.  In both images, a bright knot of
emission is also clearly present 6$\farcs$5 south of the GV Tau S
spectral trace.}
\end{figure}

The S(0) 1-0 H$_2$ emission (2.22329~$\micron$) that we detect from GV
Tau is marginally resolved in our spectra with a velocity width of 14
and 17~$\kms$ for GV Tau S and N, respectively.  Spatially, the bulk
of the emission is coincident with GV Tau N and GV Tau S (within the
0$\farcs$5 extraction aperture) but extends along the slit to the
south (P.A.=176$\arcdeg$) ending at a bright knot 6$\farcs$5 south of
GV Tau S (Fig. \ref{fig-gvtau.h2emis}).

HST imaging of the GV Tau system shows resolved nebular emission
extending away in an arc to the East, South and Southwest of the GV
Tau S while GV Tau N is comparatively free of nebulosity
(K. Stapelfeldt, private communication).  The location of the emission
knot is spatially coincident with a wisp of faint optical nebulosity
in the HST image.  No near-IR line emission is seen immediately to the
North of GV Tau, which is coincident with a lack of optical nebulosity
in this region.  The radial velocity of the H$_2$ emission is constant
along the slit ($\vlsr=8.5~\kms$), and agrees with the radial velocity
of the HCN absorption (\S~\ref{sec-molecular.abs}) and molecular
envelope for this system \citep{hogerheijde1998}.

In order 37, we detect S(2) 1-0 H$_2$ emission in both stellar
components.  The emission is extended southward along the slit in the
same way as the S(0) emission.  The emission is 2--3 times brighter
than the S(0) line emission.  As with the S(0) line, the S(2)
equivalent width is $\sim 2$ times stronger in GV Tau S compared to GV
Tau N.  In both components, the equivalent widths and velocity widths
are greater than their counterparts in the S(0) lines.  The observed
H$_2$ line ratios (i.e., 1-0 S(2)/1-0 S(0)) in GV Tau N and S are
consistent with shock excitation, similar to most of the classical T
Tauri stars observed by \citet{beck2007}.

\subsection{Molecular Absorption}\label{sec-molecular.abs}

\begin{figure}
\epsscale{1}
\plotone{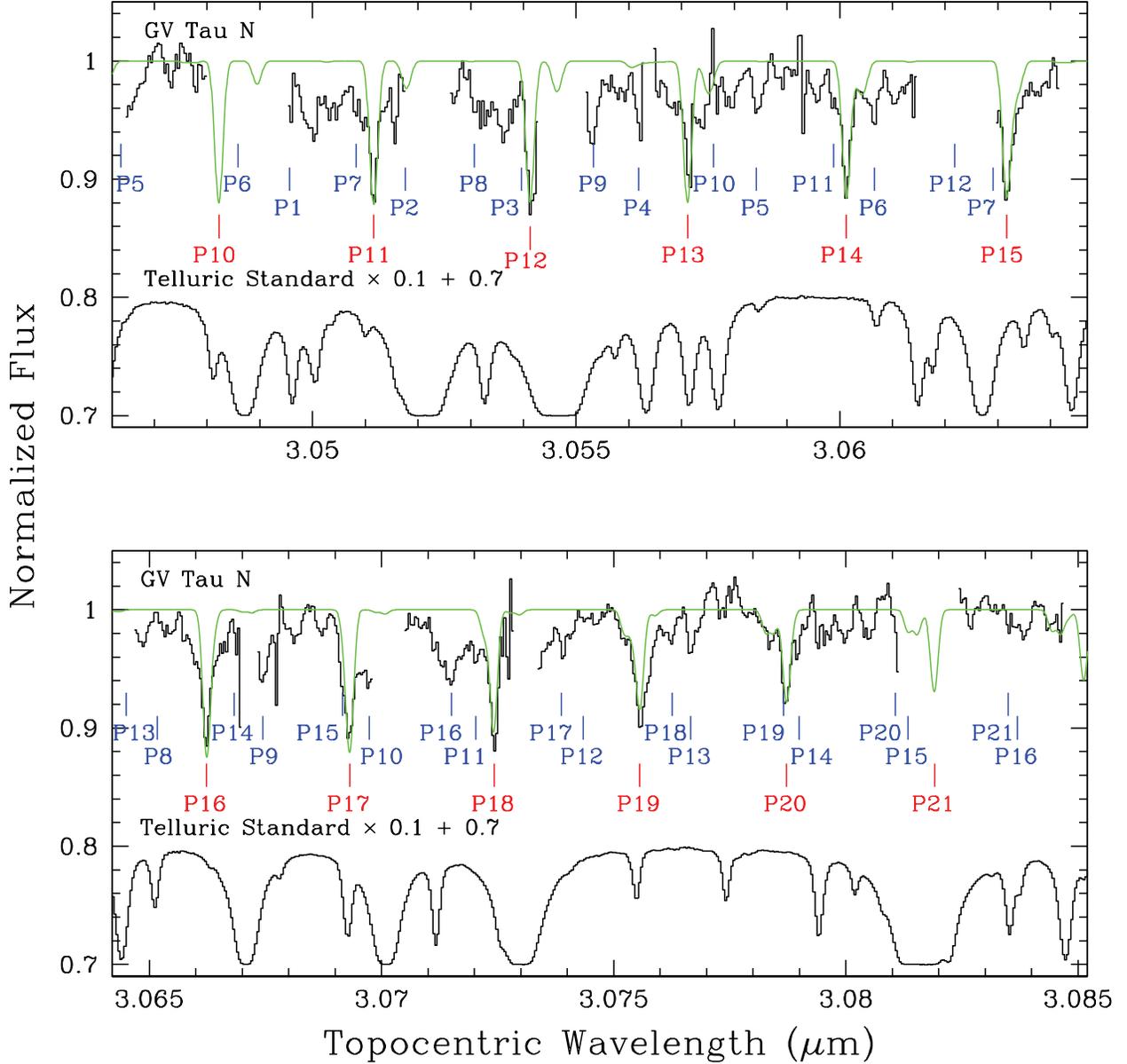}
\caption
{\label{fig-lband.gvtn.modelhcn}
$L$-band spectra of GV Tau N and the telluric standard, HR 1412 (black
histograms) in the topocentric velocity frame where the observed HCN
lines (red tickmarks) are redshifted by $35.3\kms$.  Spectral regions
with poor telluric cancelation have been excised from the GV Tau N
spectrum.  The blue tickmarks denote the expected positions of
$C_2H_2$ lines ($\nu _3$ and $\nu _2 +(\nu _4 + \nu _5)^0_+$, lower
and upper sets, respectively).  A synthetic model of HCN absorption in
a disk with temperature $T$ = 550 K and column density $N=1.5\times
10^{17}$cm$^{-2}$ (green line) provides a good fit to the
observations.}
\end{figure}

\begin{figure}
\vspace*{-17mm}
\plotone{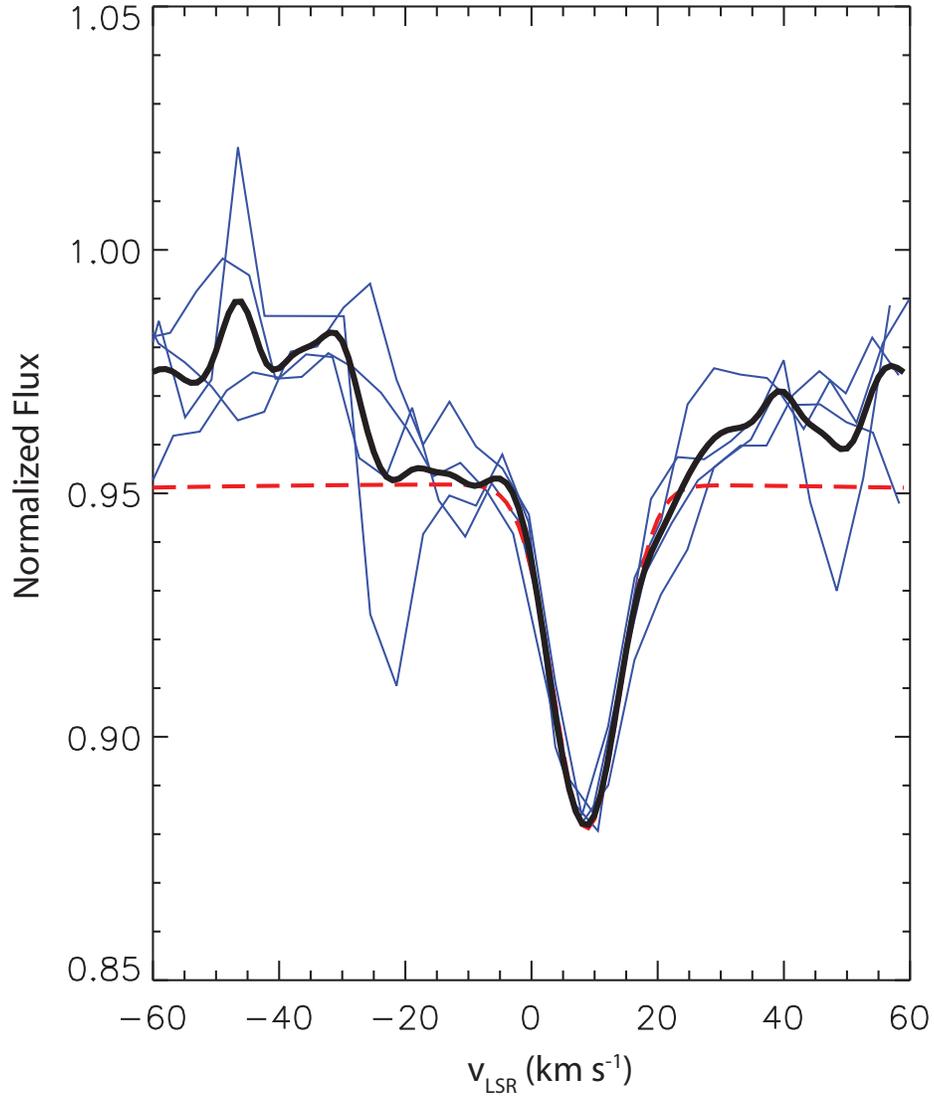}
\caption
{\label{fig-gvtn.hcnprofile}
HCN absorption lines in GV Tau N in spectral regions with atmospheric
transmission $\ge$ 70\% (P11, P14, P15, and P16 transitions, thin blue
lines) averaged together (thick black line) and fit with a Gaussian
profile (dashed red).  The velocity centroid and 1-$\sigma$ error of
the Gaussian fit to the average profile is $8.7 \pm 0.3~\kms$
($\vlsr$).}
\end{figure}

In order to characterize the molecular absorption detected in the
$L$-band spectrum of GV Tau N, we used a simple model of absorption by
a slab at a single temperature and column density.  We adopted the HCN linelist from the HITRAN database \citep{rothman1998}.
The data are well fit with a slab temperature $T=550~$K and column
density $N_{\rm HCN}=1.5 \times 10^{17} \rm{cm}^{-2}$, with a
microturbulent line broadening of $v_{\rm turb}=3~\kms$ and radial
velocity $\vlsr = 8.7 \pm 1.0~\kms$
(Fig. \ref{fig-lband.gvtn.modelhcn}).  The radial velocity of the
molecular absorption was measured using selected P-branch lines (Table
3) in regions of the spectrum that had good telluric transmission and
that were unblended with other species (such as C$_2$H$_2$).  We
constructed an average HCN absorption profile from four selected HCN
absorption transitions (P11, P14, P15, and P16) to which we fit a
Gaussian profile (Fig. \ref{fig-gvtn.hcnprofile}).  The 3-$\sigma$
error of $1.0 \kms$ on the radial velocity was estimated by comparing
the Gaussian fit of the average absorption profile to each of the
P-branch line fits separately.

\section{Discussion}

To summarize, both GV Tau N and S appear to possess slowly rotating
($\vsini=15-24\kms$) late type ($\sim$K7--M2) photospheres.  While 
the stellar properties are consistent with an age of $\sim 3$\, Myr based on current pre-main-sequence stellar evolutionary tracks, the age estimate is uncertain because ages inferred from the tracks have 
yet to be observationally verified.  We also find that GV Tau S is a 
radial velocity variable.  For GV Tau N, we
find that the system shows strong CO overtone emission, strong
$K$-band veiling ($r_K=12-15$), and a blue-shifted Br$\gamma$
centroid.  These properties are consistent with a young star
undergoing active accretion in a disk and stellar magnetosphere.  The
stellar radial velocity found for GV Tau N ($\vlsr=-4.5 \pm4.0~\kms$)
differs from the radial velocity of the GV Tau molecular envelope
\citep[$\vlsr=7.0\pm0.5~\kms$,][]{hogerheijde1998} and the radial
velocity of the warm ($\sim 550$~K) HCN absorption detected in the
$L$-band ($\vlsr=8.7 \pm 1.0\kms$).  We discuss the possible
implications of these results in the rest of this section.

\subsection{Radial Velocity Variability in GV Tau S}\label{sec-gvts.binary}

The stellar radial velocity of $\vlsr = 9.4 \pm1.7~\kms$ found for GV Tau S
differs from the $\vlsr$ observed in November 2001
\citep[$-6.2 \pm1.5~\kms$;][]{covey2006}, and the $\vlsr$ in December
1999 \citep[$3.1 \pm4~\kms$, see Table 4;][]{white2004}.  The $\vlsr$
reported here differs slightly from the $6.2 \pm0.5\kms$ velocity of
the Taurus molecular cloud in the vicinity of GV Tau, as measured in
the CO $J$=1-0 transition \citep{dame2001} and the $\vlsr= 7.0
\pm0.5~\kms$ measured for the gaseous envelope surrounding GV Tau
\citep{hogerheijde1998}.

A plausible explanation for the radial velocity variations exhibited
by GV Tau S is that it is a spectroscopic binary with a secondary that
is sufficiently faint that it is undetectable in our spectra.
Assuming that the systemic velocity is that of the molecular cloud and
that the primary has the stellar properties ($\teff,M_*$) derived
earlier (\S~\ref{sec-gvts.stellprops}), the three reported radial
velocities for GV Tau S can be accounted for if GV Tau S possesses a
$\sim 0.13 \Msun$ companion in a $P_{\rm{orbit}}\sim 38$ day circular
orbit ($i=90, a=0.35$\,AU).  A companion with this mass at the typical
age of Taurus sources ($\sim1-3$ Myr), would contribute a small
fraction (20\%) of the observed $K$ continuum.  Its $K$-band spectral
features would appear only $\sim$2\% deep in the observed spectrum,
consistent with the apparent lack of such features in our observed
spectrum.  The observed radial velocity variation and the $K$-band
spectrum are also consistent with smaller companion masses and orbital
separations. A larger stellar mass for the companion in a more face-on
orbit is ruled out since it would be detectable in our high
signal-to-noise spectra.

\subsection{Warm Molecular Absorption in GV Tau N}\label{sec-gvtn.mol.abs}

\citet{gibb2007} previously reported absorption toward GV Tau S in the
HCN low-$J$ R and P lines near $3.0~\micron$, based on observations
made on 2006 February 17 and 18.  The HCN absorption was characterized by
equivalent widths $\sim0.015~\invcm$, a radial velocity
$\vlsr=9.2\pm1.9~\kms$, and a rotational temperature of $T=115$~K.  No
HCN absorption was reported toward GV Tau N in the \citet{gibb2007}
study.  The wavelength coverage of our observations includes many of
the same lines that were observed by \citet{gibb2007} as well as
higher rotational lines.
 
At the epoch of our observations, no HCN absorption was present toward
GV Tau S.  However, HCN absorption was detected toward GV Tau N.  The
absorption lines we detect toward GV Tau N have equivalent widths and
radial velocities similar to those reported by \citet{gibb2007} toward
GV Tau S.

We find a higher temperature ($\sim550$~K) for the absorbing gas
toward GV Tau N, but this temperature is consistent with the relative
HCN line strengths observed by \citet{gibb2007}.  Their lower
temperature was based on analysis of only low-$J$ lines (i.e., R1--R6,
P2--P7); warmer temperatures are indicated when the higher-$J$ lines
used in our analysis are included.  Consistent with our results, HCN
absorption with properties similar to those reported here was also
detected toward GV Tau N (but not GV Tau S) in mid-infrared
observations made with TEXES on Gemini-North at an intermediate epoch
(November 2006; Najita et al., in preparation).

These results may indicate that the molecular absorption toward both
components is variable.  Indeed, both GV Tau N and S are found to
experience significant photometric variability in the $H$- through
$M$-bands and in the $3.1~\micron$ ice feature (Leinert et al.\ 2001).
The high temperature of the absorbing gas suggests that the gas is
located close to the star rather than in a distant circumbinary
envelope or the molecular cloud.  So if variability accounts for the
difference in the absorption properties, we are likely to be observing
the variations in the near-circumstellar environment of each source,
rather than the motion of a distant absorber across our
line-of-sight. Such a distant absorber is believed to account for the
photometric variations in GV Tau S and perhaps variations in the ice
absorption \citep{leinert2001}.

Another, more mundane, possibility is that the components of the
binary pair were misidentified by \citet{gibb2007}.  The
interpretation that the warm molecular absorption is entirely
associated with GV Tau N would be consistent with the systematically
stronger $3.1~\micron$ ice feature measured toward this component, the
strong silicate absorption that is detected only toward GV Tau N (Van
Cleve et al.\ 1994), and the hypothesis from the literature that GV
Tau N is surrounded by an edge-on disk.  Additional observations of
the GV Tau system would be useful in resolving this issue.

Setting aside for now the possibility that the HCN absorption is time
variable, we can explore the nature of the absorbing gas that we
observed based on the stellar and circumstellar properties that we
have found for GV Tau N.  As described at the beginning of this
section, the stellar velocity is blueshifted from the envelope
velocity by $\sim 11~\kms$.  This suggests that the detected stellar
component of GV Tau N possesses an orbiting companion.  If it does
not, at a velocity of $ 11~\kms$ relative to the envelope, the star
would travel 0.1 pc in $10^4$\,yr, escaping its molecular envelope on
a timescale much less than the age of the system.  If GV Tau N is then
a binary, the envelope velocity is the more appropriate velocity
reference frame for the system.  Another possibility is that the
stellar light is not seen directly, but rather in reflection against
material on the far side of the envelope that is infalling toward us
(S. Strom, private communication).  In either interpretation, the
measured stellar velocity is not the systemic velocity.

As also described at the beginning of this section, the radial
velocity of the HCN absorption is marginally different from that of
the GV Tau molecular envelope by $1.7\pm1.0~\kms$.  The small velocity
difference allows for the possibility that some of the absorption
arises in gas that is infalling with respect to the envelope.  Higher
spectral resolution observations (e.g., with TEXES; Najita et al., in
preparation) can better determine whether this is the case.

In the meantime, we can consider two possible scenarios for the origin
of the HCN absorption: (1) Some fraction of the absorption arises in
gas infalling from the molecular envelope toward the star at $\sim
2~\kms$ and/or (2) the absorption arises in warm gas in orbit around
the star.

{\it Origin in an Envelope.} Since the bolometric luminosity of
the system is 7--9$\Lsun$ \citep{kenyon1995, furlan2007} and the
stellar contribution (photospheric and stellar accretion,
\S~\ref{sec-gvtn.stellprops} and \S~\ref{sec-brg.emis}) to the
luminosity is $\sim1.0\Lsun$, the residual disk and envelope accretion
is $L_{\rm acc} \sim6-8\Lsun$.  The large residual accretion
luminosity compared to the stellar accretion luminosity for GV Tau N
inferred from the Br$\gamma$ emission ($\sim0.2\Lsun$;
\S~\ref{sec-brg.emis}) implies that the disk/envelope accretion rate
is much larger than the stellar accretion rate for GV Tau N, similar
to the results obtained for Class I sources by \citet{white2004}.  For
accretion in a disk inward to a radius of $R_{\rm{in}}$, the disk
accretion luminosity $L_d=GM_*\Mdotdisk / 2R_{\rm{in}}$ of 7$\Lsun$
corresponds to a disk accretion rate of $\Mdotdisk \sim10^{-6}\Msunperyr$
for $R_{\rm{in}}=2.5 \Rsun$ and a stellar mass of $\Mstar = 0.8\Msun$.
In comparison, the spectral energy distributions of most of the Class
I sources in Taurus are fit with envelope accretion rates up to an
order of magnitude larger \citep{furlan2007}.  A lower accretion rate
$\sim 10^{-6}\Msunperyr$ is more consistent with the weak envelope
emission and outflow activity associated with GV Tau.

For the estimated mass of the stellar component(s) of GV Tau N
(0.8$\Msun$), the envelope infall velocity $v_r = (2GM_*/r)^{1/2}$ is
expected to be $\sim 2~\kms$ at distances of 360\,AU.  In contrast,
models of gas in collapsing protostellar envelopes suggest that the
observed HCN temperature of $\sim 500$~K can be achieved only close to
the star, within 2\,AU given the accretion luminosity of the system
\citep[][their figure 4]{ceccarelli1996}, i.e., well within the
distance of 360\,AU that would be inferred for the infalling gas based
on its velocity relative to the molecular envelope.  It therefore
appears unlikely that most of the absorption could arise in an
infalling envelope unless there are additional heating processes for
the infalling gas beyond those considered in
\citet{ceccarelli1996}.  
If other heating processes, such as oblique shocks, can heat distant
material to $\sim 500$\,K, the warm molecular absorption that we
observe could also arise in a distant, non-infalling envelope.  However, 
the lack of molecular absorption observed toward GV Tau S argues
against the gas originating from a distant envelope which would surround both GV Tau N and S.

An alternative scenario is that GV Tau N is a single star and its
measured stellar velocity is the appropriate systemic velocity.  In
this case, at a distance 2\,AU from the star, the radial component of
the infall velocity would be $(G\Mstar/r)^{1/2} \sim $20~$\kms$
\citep{cassen1981} for the estimated stellar mass of GV Tau N.  This
is similar to the $13~\kms$ velocity difference between the HCN
absorption and GV Tau N stellar photosphere (Table 4).  If GV Tau N is
a single star, it is then moving relative to the molecular envelope
and, as already discussed above, it will travel $\sim 0.1$\,pc in
$<10^4$\,yr.  It then seems unlikely that the star and the envelope
are physically associated, which makes it difficult to account for the
origin of the infalling gas in this scenario.

{\it Origin in a Circumstellar Disk.}  Another possibility is that
the HCN absorption arises in the heated atmosphere of a gaseous disk
\citep[see also][]{gibb2007}.  Models of the chemistry of the inner
regions of circumstellar disks predict that HCN will be abundant at
radial distances of a few AU \citep{markwick2002}.  Models of the
thermal structure of disk atmospheres further predict that the disk
surface will reach temperatures $\gtrsim 500$\,K at distances of a few
AU \citep[e.g.,][]{glassgold2004}.  If the disk in the GV Tau N system
is viewed edge-on, as hypothesized in the literature
\citep{menard1993,vancleve1994}, it seems plausible that the warm HCN
absorption could arise in the disk atmosphere seen in absorption
against the warmer, inner region of the disk that produces the
$L$-band continuum.

The rarity of such a line--of--sight through a disk may also explain
another unusual characteristic of GV Tau: the low CO$_2$ ice
absorption optical depth toward the source compared to its silicate
optical depth.  The data of \citet{furlan2007} show that the peak
10~$\micron$ silicate optical depth is $\sim9$ times the peak optical
depth of the 15.2~$\micron$ CO$_2$ ice absorption.  All other Taurus
Class I sources in the Furlan sample have $\tau_{\rm
silicate}/\tau_{\rm CO_2} \lesssim 4$, with the exception of DG Tau B,
another possible edge-on source.  A high ratio of $\tau_{\rm
silicate}/\tau_{\rm CO_2}$ may occur if a substantial fraction of the
silicate optical depth is produced in a warm disk region in which
CO$_2$ ice has sublimated.

GV Tau N would be a spectroscopic binary in this scenario, in order to
account for the difference in radial velocity between the stellar
photosphere and the warm molecular absorption.  Because the warm HCN
absorption is approximately at rest relative to the molecular envelope
velocity, in this scenario it would arise in a {\it circumbinary} disk
seen edge-on.  In contrast, the CO overtone emission, which is found
to share the radial velocity of the stellar photosphere, arises from
the {\it circumstellar} disk of the primary component.

\subsection{Comparison to IRS 46}

Strong absorption by gaseous warm organic molecules appears to occur
rarely among low mass young stars.  GV Tau and another source, IRS 46,
are the only systems in which such absorption has been reported to
date.  IRS 46 is a low-mass YSO with a spectral energy distribution
that indicates either a Class I source or a Class II source viewed
nearly edge-on \citep{lahuis2006}.  IRS 46 is similar to GV Tau N in
that it also shows Br$\gamma$ and 2.3~$\micron$ CO overtone emission \citep[see YLW 16B in][]{doppmann2005} and strong absorption by gaseous warm organic
molecules in the $L$-band and at mid-infared wavelengths \citep[CO,
HCN, C$_2$H$_2$, CO$_2$;][]{lahuis2006}.  The temperature
($\sim400$~K) and equivalent width of the HCN absorption are similar
to that observed toward GV Tau N.  Unlike GV Tau N, the warm HCN
absorption in IRS 46 ($\vlsr = -20~\kms$) shows a large velocity
offset from the surrounding molecular cloud ($\vlsr = 4.4~\kms$, Table
4).  In contrast to the situation for GV Tau N, the stellar radial
velocity of IRS 46 is unknown.  If the system is at the cloud
velocity, the warm molecular absorption is {\it blueshifted} from the
system by $24~\kms$.  \citet{lahuis2006} speculated that the warm
molecular absorption in IRS 46 arises in the inner region of a
circumstellar disk, possibly at the footpoint of an outflowing disk
wind in order to account for the presumed blueshift of the absorption.
This scenario is consistent with the interpretation of the spectral
energy distribution as indicating a disk system viewed edge-on.  Our
``disk origin'' scenario for GV Tau N is similar to the explanation
given by \citet{lahuis2006} for the HCN absorption in IRS 46.  In both
cases, the warm molecular absorption arises in a disk,
although there is no outflow component to the flow in the case of GV
Tau N.

\section{Summary and Conclusions}

We have used spatially resolved $K$- and $L$-band spectroscopy to
characterize the stellar and circumstellar properties of the GV Tau
binary system.  We find that GV Tau S is a radial velocity variable,
possibly the result of an unseen low-mass companion.  The radial
velocities measured here and in the literature, when combined with the
apparent absence of the spectral signature of a companion in our
$K$-band spectra, are consistent with a companion with $M_*< 0.15
\Msun$ and $a< 0.35$\,AU.  Further spectroscopic monitoring of this
source would be useful to confirm our interpretation and to better
constrain the companion properties.

The other component of the binary system, GV Tau N, is found to show
strong CO overtone emission, strong $K$-band veiling ($\rk=12-15$),
and Br$\gamma$ emission with a blue-shifted emission centroid;
signatures consistent with a young actively accreting star.  From the
presence of apparent stellar absorption features in the 2-0 CO
overtone bandhead and the 2.2~$\micron$ Na emission features, we infer
a late spectral type ($\teff \sim 4100$~K) and a stellar radial
velocity of $\vlsr = -4.5~\kms$.  The warm ($T=550$\,K) HCN absorption
found toward GV Tau N in the $L$-band spectrum of this source is
offset in velocity by $\sim 13~\kms$ relative to the star and $\sim
2~\kms$ relative to the GV Tau molecular envelope.  The large velocity
of the star relative to its molecular envelope would cause it to
escape the molecular envelope on a timescale much shorter than its age
unless the star is a spectroscopic binary.

The small redshift ($1.7\pm 1.0~\kms$) of the warm molecular
absorption relative to the molecular envelope velocity (assumed to be
the systemic velocity) suggests that most of the absorption arises in a disk atmosphere viewed close to edge on, although an
origin in the molecular envelope is not ruled out completely.  This
interpretation can be tested with further observations of GV Tau N.
Observations at higher spectral resolution can better constrain the
radial velocity of the absorbing gas relative to the systemic
velocity.  Observations that explore the molecular abundances of the
absorbing gas may also be useful in testing a disk origin for the
absorption.

\acknowledgments

The authors wish to recognize and acknowledge the very significant
cultural role and reverence that the summit of Mauna Kea has always
had within the indigenous Hawaiian community.  We are most fortunate
to have the opportunity to conduct observations from this mountain.
We thank Al Conrad and other Keck Observatory staff who provided
support and assistance during our NIRSPEC run.  We thank Sean Brittain
for sharing results from \citet{gibb2007} in advance of publication.
We thank Elise Furlan for information on the silicate and CO$_2$ ice
optical depths of Taurus Class I sources in advance of publication.
Nathan Crockett contributed helpful advice on data-reduction
strategies.  Financial support for this work was provided by the NASA
Origins of Solar Systems program and the NASA Astrobiology Institute
under Cooperative Agreement No.\ CAN-02-OSS-02 issued through the
Office of Space Science.  This work was also supported by the Life and
Planets Astrobiology Center (LAPLACE).



{\it Facility:} \facility{Keck:II(NIRSPEC)}.

\clearpage



\begin{deluxetable}{cccccl}
\tablecolumns{6}
\tablewidth{-1pt}
\tablecaption{GV Tau South: Radial Velocity Measurements}
\tablehead{
\colhead{Order/Wavelength} &
\colhead{Lines(s)}   & 
\colhead{absorption}    & 
\colhead{emission} &
\colhead{$\vlsr$}   & 
\colhead{notes}\\
&
&
&
&
\colhead{($\kms$)}}

\startdata

Order 37, 2.03~$\micron$ &	S(2) 1-0 H$_2$ &	no &	yes &	10.3  & line centroid\\
Order 36, 2.11~$\micron$ &	Mg I, Al I&	yes &	no &	...  & radial velocity poorly constrained in this region\\
Order 35, 2.16~$\micron$ &	Br$\gamma$ &	no &	yes &	3.7  & line centroid\\
Order 34, 2.21~$\micron$ &	Na I, Ti I, Fe I &	yes &	no &	9.1$\pm$1.3  & mean value of 7 lines\\
Order 34, 2.22~$\micron$ &    S(0) 1-0 H$_2$  &	 no &	 yes &	8.5 & emission extends south of the star\\
Order 33, 2.29~$\micron$ &  Fe I, Ca I, Ti I &	yes &	no &	10.0$\pm$1.3  & mean value of 4 lines\\
Order 33, 2.29~$\micron$ &  $\nu$=2-0 CO  &	yes &	yes &	...  & radial velocity poorly constrained in this region\\
Order 25, 3.07~$\micron$ &	HCN, C$_2$H$_2$ &	no &	no & ...  & \\

\enddata
\end{deluxetable}



\begin{deluxetable}{cccccl}
\tablecolumns{6}
\tablewidth{-1pt}
\tablecaption{GV Tau North: Radial Velocity Measurements}
\tablehead{
\colhead{Order/Wavelength} &
\colhead{Lines(s)}   & 
\colhead{absorption}    & 
\colhead{emission} &
\colhead{$\vlsr$}   & 
\colhead{notes}\\
&
&
&
&
\colhead{($\kms$)}}

\startdata

Order 37, 2.03~$\micron$ &	S(2) 1-0 H$_2$ &	no &	yes &	5.9  & line centroid\\
Order 36, 2.11~$\micron$ &	Mg I, Al I &	no &	no &	  ... & \\
Order 35, 2.16~$\micron$ &	Br$\gamma$ &	no &	yes &	-17.0  & line centroid\\
Order 34, 2.21~$\micron$ &	Na I, Si I, Sc I &	yes &	yes &	-4.5$\pm$4.0  & velocity fit to Na absorption cores\\
Order 34, 2.22~$\micron$ &    S(0) 1-0 H$_2$  &	 no &	 yes &	8.5 & line centroid\\
Order 33, 2.29~$\micron$ &	$\nu$=2-0 CO  &	yes &	yes &	-4.5$\pm$4.0  & velocity fit to CO absorption core\\
Order 25, 3.07~$\micron$ &	HCN, C$_2$H$_2$ &	yes &	no &	8.7$\pm$0.3  & line centroid\\

\enddata
\end{deluxetable}



\begin{deluxetable}{cccccc}
\tablecolumns{5}
\tablewidth{-1pt}
\tablecaption{HCN Lines in GV Tau N}
\tablehead{
\colhead{Line ID} &
\colhead{$\lambda _{rest}$}   & 
\colhead{$\lambda _{topo}$}   & 
\colhead{$\vlsr$}    & 
\colhead{\%~Transmission} &
\colhead{Equivalent Width} \\
&
\colhead{\micron}& 
\colhead{\micron}&
\colhead{$\kms$}&
&
\colhead{\AA }}

\startdata

P11 & 3.0508033 & 3.0511622 & 8.73 & 0.75 &  0.12 \\
P12 & 3.0537670 & ... & ... & 0.35 & 0.21 \\
P13 & 3.0567549 & ... & ... & 0.20& 0.12 \\
P14 & 3.0597676 & 3.0601268 & 8.65 & 0.98 & 0.14 \\
P15 & 3.0628043 & 3.0631660 & 8.86 & 0.70 & 0.15 \\
P16 & 3.0658660 & 3.0662214 & 8.21 & 0.85 & 0.19 \\
P17 & 3.0689521 & ... & ... & 0.25 & 0.24 \\
P18 & 3.0720633 & ... & ... & 0.45 & 0.15 \\
P19 & 3.0751988 & ... & ... & 0.65 & 0.20 \\
P20 & 3.0783594 & 3.0787095 & 7.56 & 0.95 & 0.14 \\
P11,14,15,16 &  & & 8.7$\pm0.3$ & & \\

\enddata
\end{deluxetable}



\begin{deluxetable}{llcl}
\tablecolumns{4}
\tablewidth{-1pt}
\tablecaption{Radial Velocities for GV Tau and IRS 46}
\tablehead{
\colhead{System} &
\colhead{Component}   & 
\colhead{$v_{\rm LSR}$}   & 
\colhead{Reference} \\
&
&
\colhead{$\kms$}& 
}
\startdata
GV Tau   & Cloud	&  6.2 $\pm$0.5		& from \citet{covey2006} \\
	 & Envelope	&  7.0 $\pm$0.5		& Hogerheijde et al. (1998) \\
GV Tau S 
         & Star 	&  3.1 $\pm$3.8			& White \& Hillenbrand (2004)\\
	 &     		& -6.2 $\pm$1.5 			& Covey et al.\ (2006) \\
         &		&  9.4 $\pm$1.7			& this paper \\
GV Tau N & Star		& -4.5 $\pm$4.0	& this paper \\
	 & HCN warm abs	& 8.7 $\pm$0.3		& this paper \\
IRS46	 & Envelope	&  4.4			& Lahuis et al.\ (2006) \\
	 & CO, HCN warm abs	& -20		& Lahuis et al.\ (2006) \\
	 & Star 	& unknown	& \\
\enddata
\end{deluxetable}


\clearpage





\clearpage
{\plotone{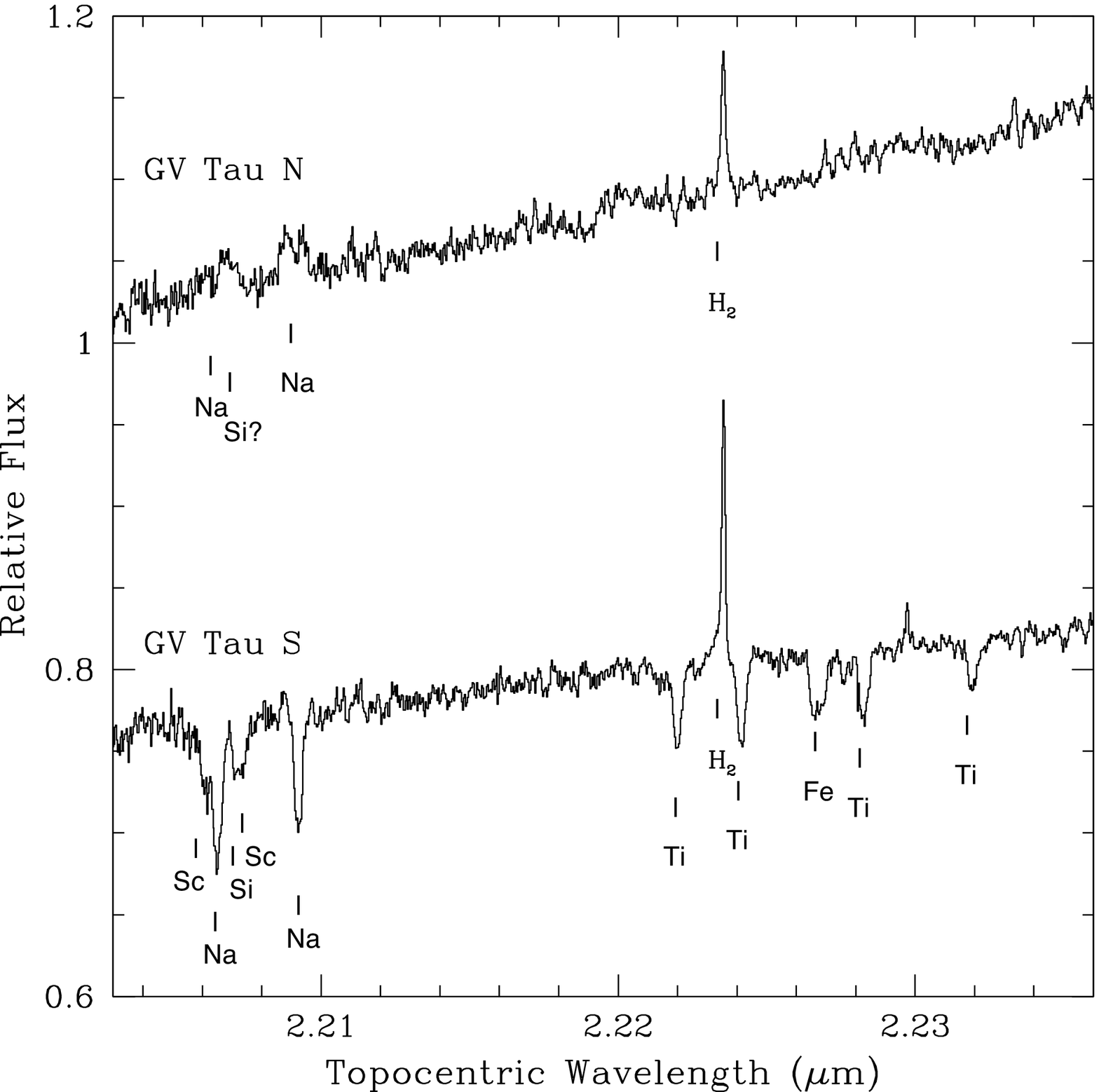}}
\centerline{Fig. 1b. --- Continued.}
\clearpage
{\plotone{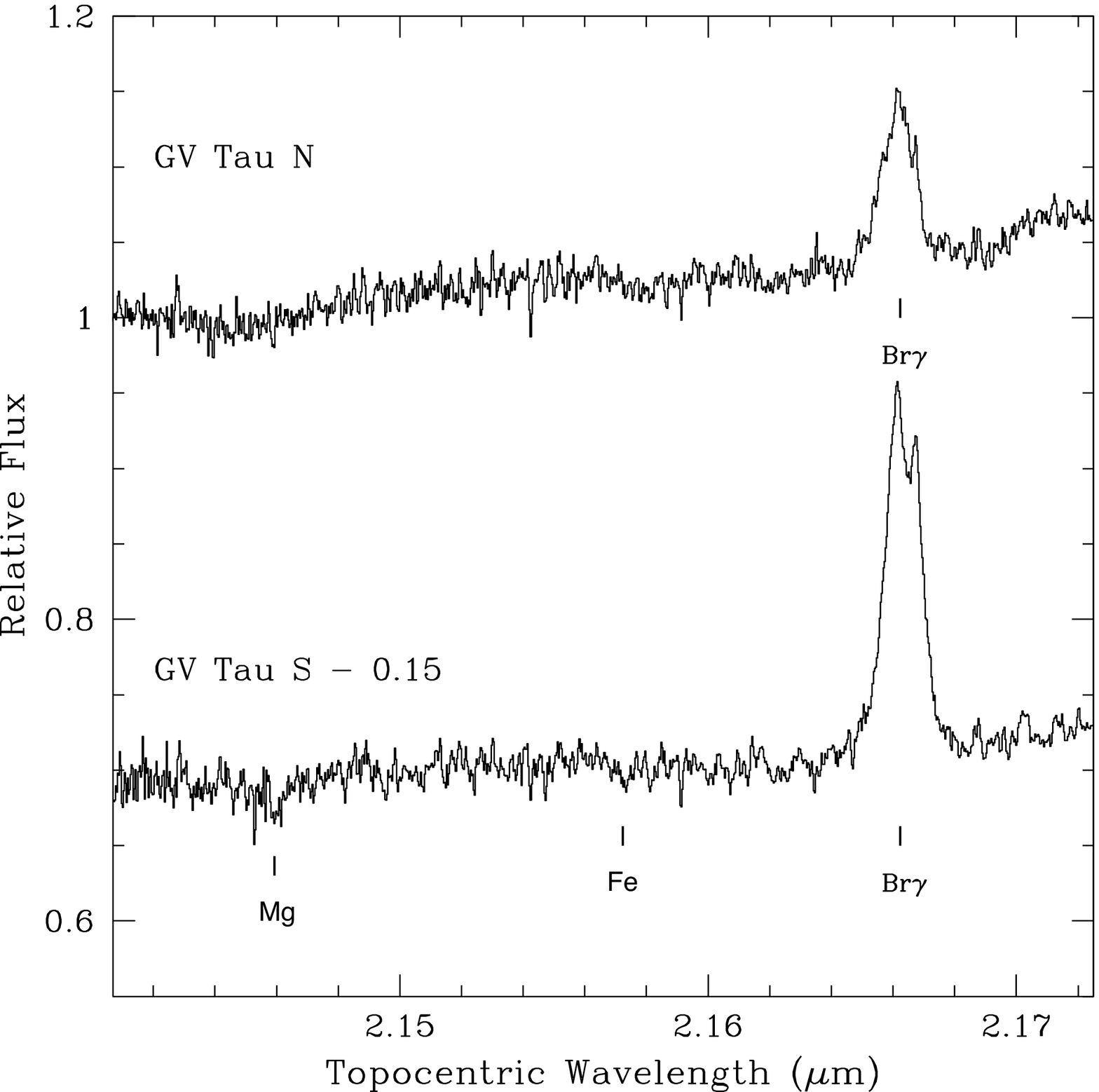}}
\centerline{Fig. 1c. --- Continued.}
\clearpage
{\plotone{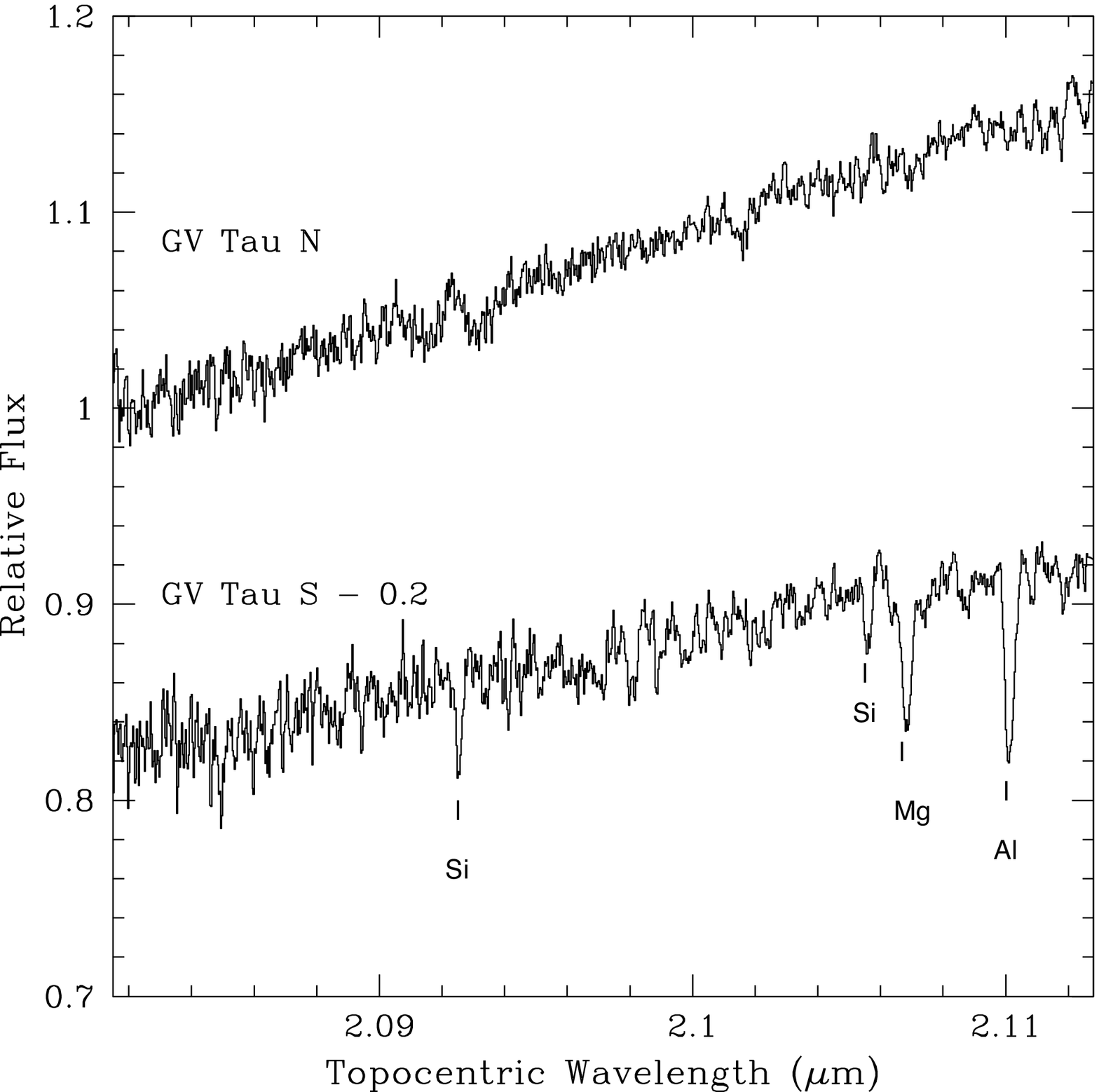}}
\centerline{Fig. 1d. --- Continued.}
\clearpage
{\plotone{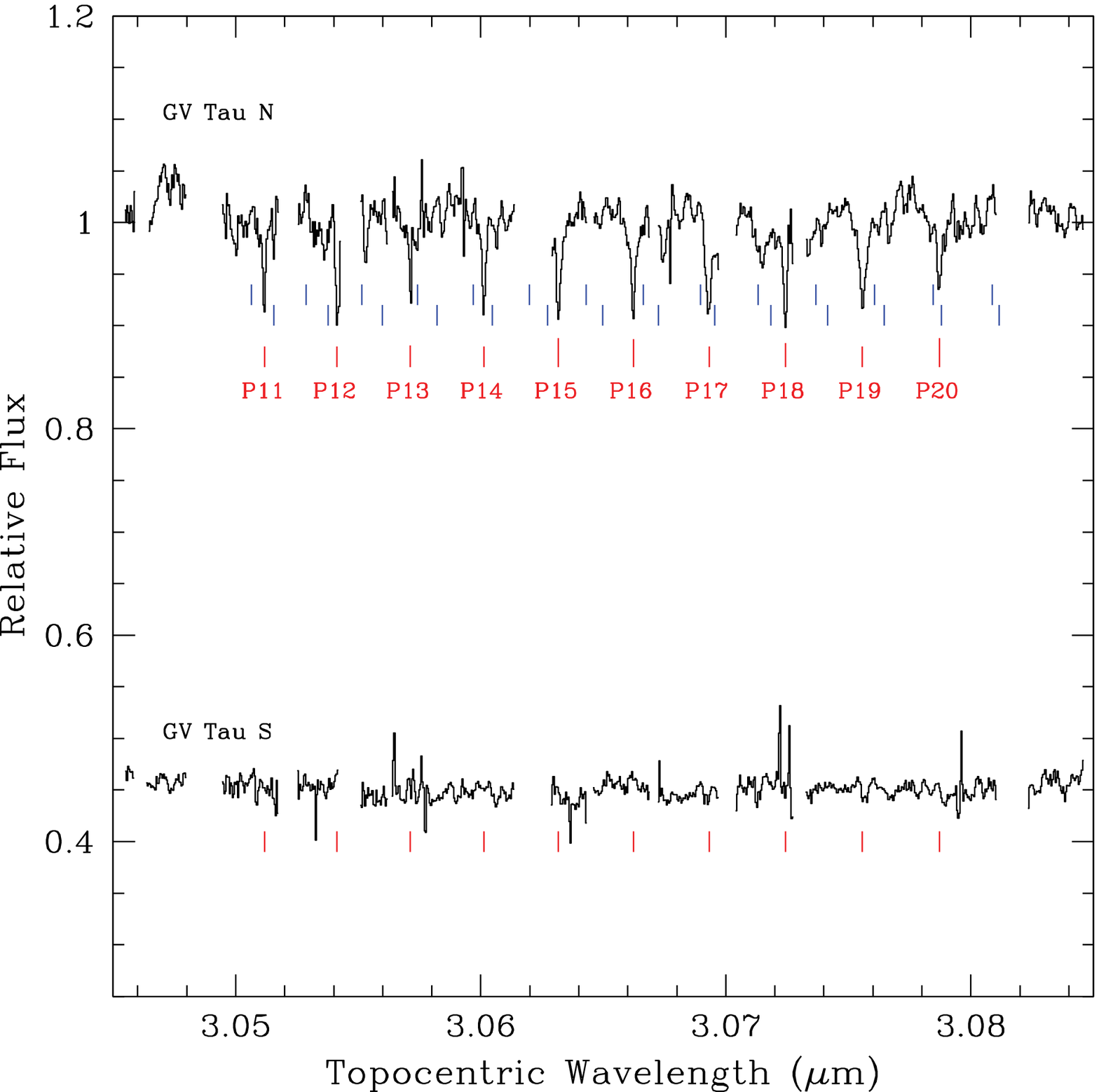}}
\centerline{Fig. 1e. --- Continued.}

\clearpage

\end{document}